\documentclass[%
 aip,
 amsmath,amssymb,
 preprint,%
]{revtex4-1}

\usepackage{graphicx}
\usepackage{dcolumn}
\usepackage{bm}
\usepackage{float}
\makeatletter
\let\newfloat\newfloat@ltx
\makeatother
\usepackage{algorithm}
\usepackage{algpseudocode}
 \usepackage{hyperref}
\usepackage[T1]{fontenc}
\usepackage{mathptmx}
\usepackage[section]{placeins}
\usepackage{tikz}
\usepackage{etoolbox}
\usepackage{array}
\makeatletter
\patchcmd{\chapter}{\if@openright\cleardoublepage\else\clearpage\fi}{}{}{}
\makeatother
\usepackage{graphicx}
\usepackage{balance}
\usepackage{tablefootnote}
\usepackage{soul}
\hypersetup{
    colorlinks,
    linkcolor={red!50!black},
    citecolor={blue!50!black},
    urlcolor={blue!80!black}
}
\tikzset{neuron/.style={shape=circle, minimum size=2cm, inner sep=0.2, draw, font=\large}, 
  input_io/.style={shape=circle, minimum size=2.0cm, inner sep=0.2, draw, font=\large, fill=cyan!10}, 
  output_io/.style={shape=circle, minimum size=2.0cm, inner sep=0.2, draw, font=\large, fill=orange!15}, 
  dense/.style={shape=ellipse, minimum width=1.75cm, minimum height=0.8cm, inner sep=0.2, draw, font=\large}}
  \tikzset{font=\normalsize}
 \usetikzlibrary{positioning, fit, arrows.meta, shapes}
\makeatletter
\def\@email#1#2{%
 \endgroup
 \patchcmd{\titleblock@produce}
  {\frontmatter@RRAPformat}
  {\frontmatter@RRAPformat{\produce@RRAP{*#1\href{mailto:#2}{#2}}}\frontmatter@RRAPformat}
  {}{}
}%

\makeatother

\def\bv{{\mathbf{v}}}
\def\bV{{\boldsymbol{V}}}
\def\bM{{\boldsymbol{M}}}
\def\bD{{\mathbf{D}}}
\def\bC{{\boldsymbol{C}}}
\def\bJ{{\boldsymbol{J}}}
\def\bQ{{\boldsymbol{Q}}}
\def\bR{{\boldsymbol{R}}}

\begin{document}

\preprint{AIP/123-QED}

\title[Reconstruction, forecasting, and stability of chaotic dynamics from partial data]{Reconstruction, forecasting, and stability of chaotic dynamics from partial data}
\author{Elise Özalp}
\author{Georgios Margazoglou}%
\affiliation{ 
Department of Aeronautics, Imperial College London, London SW7 2BX, United Kingdom
}%
\author{ Luca Magri}
 \email{l.magri@imperial.ac.uk.}
\affiliation{ 
Department of Aeronautics, Imperial College London, London SW7 2BX, United Kingdom
}%
 \affiliation{The Alan Turing Institute, London NW1 2DB, United Kingdom}

\date{24 May 2023}

\begin{abstract}
The forecasting and computation of the stability of chaotic systems from partial observations are tasks for which traditional equation-based methods may not be suitable. 
In this computational paper, we propose data-driven methods to
(i) infer the dynamics of unobserved (hidden) chaotic variables (full-state reconstruction);
(ii) time forecast the evolution of the full state; 
and (iii) infer the stability properties of the full state. 
The tasks are performed with long short-term memory (LSTM) networks, which are trained with observations (data) limited to only part of the state:
(i)  the low-to-high resolution LSTM (LH-LSTM), which takes partial observations as training input, and requires access to the full system state when computing the loss;
and (ii) the physics-informed LSTM (PI-LSTM), which is designed to combine partial observations with the integral formulation of the dynamical system's evolution equations.
First, we derive the Jacobian of the LSTMs. Second, 
 we analyse a chaotic partial differential equation, the Kuramoto-Sivashinsky (KS), and the Lorenz-96 system. 
We show that the proposed networks can forecast the hidden variables, both time-accurately and statistically. 
The Lyapunov exponents and covariant Lyapunov vectors, which characterize the stability of the chaotic attractors, are correctly inferred from partial observations. 
Third, the PI-LSTM outperforms the LH-LSTM by successfully reconstructing the hidden chaotic dynamics when the input dimension is smaller or similar to the Kaplan-Yorke dimension of the attractor. \textcolor{black}{The performance is also analysed against noisy data.}
This work opens new opportunities for reconstructing the full state, inferring hidden variables, and computing the stability of chaotic systems from partial data.  

\end{abstract}

\maketitle

\begin{quotation}
It has been shown that the chaotic dynamics of complex systems can be inferred by machine learning from data \textcolor{black}{when all the dynamics are observed, but questions remain when only partial observations are available}. We design data-driven approaches that can accurately infer hidden chaotic dynamics from partial data, and correctly learn the chaotic attractor, \textcolor{black}{its tangent space}, and its stability properties. 
\end{quotation}

\section{Introduction}

Chaotic dynamics appear in applications in various fields, from meteorology \cite{lorenz63}, through engineering and chemistry \cite{strogatz1994nonlinear}, to propulsion~\cite{huhn2020stability,magri2020physics}. Chaos emerges due to the system's exponential sensitivity to initial conditions, as is the case with turbulent fluid dynamics \cite{takens1981detecting}. However, the accurate prediction of chaotic spatiotemporal behaviour is challenging because numerical computations can be computationally expensive, and sensor measurements are often limited, providing only a partial observation of the dynamic behaviour. These partial observations can lead to a misrepresentation of the unobserved (hidden) variables, or other long-term physical properties, which further challenges the modelling and prediction of the systems. 

The predictability and stability of a chaotic system are characterized by its tangent space, which can be computed using the linearized dynamics provided by the Jacobian. This computation allows for the derivation of quantities such as the Lyapunov exponents (LEs), which measure the exponential rate of separation of trajectories. A geometric characterization is provided by the covariant Lyapunov vectors (CLVs), which constitute a covariant basis of the tangent space, and point to directions of asymptotic expansion and contraction of the dynamical system \cite{Ginelli2013_clv}. 
Preserving these stability properties is crucial when building surrogate models from limited observations to a more comprehensive dataset~\cite{takeuchi2011hyperbolic}. 

Neural networks are an expressive nonlinear representation of continuous functions, which can extract patterns from data and, once trained, provide computationally cheap predictions. Suitable for time series and dynamical evolutions are recurrent neural networks (RNNs), which have shown promising performance in the inference of dynamical systems with multi-scale, chaotic, or turbulent behaviour \cite{vlachas2018data}. There are several classes of RNNs, the main classes being reservoir computing, such as echo state networks (ESNs) \cite{jaeger2001echo}, and networks using backpropagation through time, such as long short-term memory networks (LSTMs) \cite{Hochreiter_1997_LongShortTermMemory}. 
\textcolor{black}{Both networks have been extensively applied to short-term prediction tasks when trained on complete observations \cite{Vlachas_2020_backprop, racca2021robust, PhysRevLett.120.024102_pathak, vlachas2018data}. Additionally, ESNs have been employed for predicting from noisy, undersampled observations and for crossprediction \cite{herzog2021reconstructing, zimmermann2018observing}, whilst LSTMs have been trained on incomplete observations \cite{ouala2020learning, brajard2020combining} with a focus on short-term time metrics such as the root-mean-square error. 
In the context of learning stability properties for prototypical chaotic systems, ESNs have demonstrated accurate inference capabilities when trained on the full state \cite{Margazoglou2023, Pathak_2017_ml_le, Vlachas_2020_backprop}.}
This work proposes two approaches to infer the hidden chaotic dynamics from partial observations (data) with LSTMs. The first approach considers the scenario in which the full-state data is only available for a limited period, and partial data is available for the remaining time. This scenario arises when generating high-resolution data is computationally expensive. For this, the goal is to analyse how well LSTMs can infer the full-state dynamics from partial data, especially in capturing small-scale statistics in partial-differential equations. The second approach analyses the case in which only partial state data is available, and the aim is to reconstruct (i.e., infer) the hidden dynamics. To achieve this, knowledge of the underlying physical system has to be incorporated by adding constraints based on the governing equations. We propose a physics-informed loss function based on an integral formulation of the dynamical system, which is versatile and suitable for the architecture of the RNN. 

\textcolor{black}{This paper has a three-fold goal. First, we perform reconstruction of the full state vector from partial data and forecast the entire state in the future. Second, we infer the Jacobian of the system, enabling the analysis of the tangent space both geometrically (with CLVs) and spectrally (with LEs and CLVs) from data only, i.e. without computing the Jacobian of the physical systems' governing equations. Lastly, both approaches are compared on partial inputs from two prototypical chaotic systems: the Lorenz-96 system \cite{lorenz96} and the Kuramoto–Sivashinsky equation \cite{Kuramoto_1978_DiffusionInducedChaos, Hyman_1986_KuramotoSivashinsky}.}

The paper is structured as follows. Section \ref{sec:LSTM} introduces the LSTM architecture to address the state reconstruction problem of a chaotic dynamical system. We propose a data-driven and physics-informed loss formulation in Section \ref{sec:loss_formulation}. In Section \ref{sec:stab_prop}, we provide an overview of the computation of LEs and CLVs, emphasizing their importance for physically accurate modelling. We derive the Jacobian of the LSTM. Section \ref{sec:results} discusses the results for the Kuramoto–Sivashinsky system, followed by the findings for the Lorenz-96 system. Finally, in Section \ref{sec:conclusion}, we summarize our work and discuss future directions.

\section{Long short-term memory}\label{sec:LSTM}
A dynamical system's solution is a time series, which provides an ordered sequence of data over time. Recurrent neural networks (RNNs) are a data-driven approach for modelling time series that utilize hidden states to encode the input's information history. Among RNNs, long short-term memory (LSTM) networks are a common tool due to their internal gating mechanisms, which mitigate the vanishing gradient problem \cite{}.

LSTMs are commonly employed for time series forecasting as they embed the time-delayed inputs in their higher-dimensional hidden states. 
From a dynamical systems perspective, a delay embedding can provide a structure that is topologically equivalent to the attractor \cite{takens1981detecting}. Given their architecture, LSTMs have the potential to be a powerful tool for representing the dynamics of a chaotic system. This enables them to be utilised for reconstructing and forecasting dynamical systems, even in cases in which we lack part of the system's full state information, as will be demonstrated in the following.

\subsection{State reconstruction for the inference of hidden dynamics}
We consider a nonlinear autonomous dynamical system
\begin{equation}\label{autonom_dyn_system}
 \frac{d}{dt} \bm{x}(t) = f(\bm{x}(t)),
\end{equation}
where $\bm{x}(t) \in \mathbb{R}^D$ is the state vector of the physical system and $f:\mathbb{R}^D \to \mathbb{R}^D$ is a smooth nonlinear vector function.  Experimental observations often come with a limited amount of information on a system's full state. For example, if in a wind tunnel experiment $\mathcal{O}(10)$ sensors are placed in different positions, they may provide adequate but scarce spatial information of the system's evolution~\cite{adrian1991particle}. Mathematically, if $\bm{x}(t) = [\bm{y}(t); \bm{\xi}(t)]$ is the full state of a chaotic dynamical system, then $\bm{y}(t) \in \mathbb{R}^{D_y}$ are the scarce observations and $\bm{\xi}(t) \in \mathbb{R}^{D_{\xi}}$ are the unobserved (hidden) variables with $D = D_y + D_{\xi}$. Specifically, let us assume that $\bm{y}(t_i)$ is measured at times $t_i = i \Delta t $ with $i=0, \dots N_{t}$ and constant time step $\Delta t$. Based on these observations, we wish to reconstruct and predict the full state $\bm{x}(t_i) =[\bm{y}(t_i), \bm{\xi}(t_i)]$. 

LSTMs have been successfully applied to time series forecasting of dynamical systems when full observations are available \cite{SANGIORGIO2021111570, vlachas2018data, Vlachas_2020_backprop}. In this work, we use the LSTM as a model of partially observed chaotic time series to both reconstruct the full state and perform accurate autonomous temporal evolution (forecasting).

\subsection{Architecture}
\begin{figure}[h]
    \centering
            \scalebox{0.7}{
\begin{tikzpicture}[
    font=\sf \scriptsize,
    >=LaTeX,
    cell/.style={
        rectangle, 
        rounded corners=5mm, 
        draw,
        very thick,
        },
    operator/.style={
        circle,
        draw,
        inner sep=-0.5pt,
        minimum height =.2cm,
        },
    function/.style={
        ellipse,
        draw,
        inner sep=1pt
        },
    ct/.style={
        circle,
        draw,
        line width = .75pt,
        minimum width=1cm,
        inner sep=1pt,
        },
    gt/.style={
        rectangle,
        draw,
        minimum width=4mm,
        minimum height=3mm,
        inner sep=1pt
        },
    mylabel/.style={
        font=\scriptsize\sffamily
        },
    ArrowC1/.style={
        rounded corners=.25cm,
        thick,
        },
    ArrowC2/.style={
        rounded corners=.5cm,
        thick,
        },
    ]

    \node [cell,  fill=gray!10, minimum height =4cm, minimum width=6cm] at (0,0) {} ;

    \node [gt] (ibox1) at (-2,-0.75) {$\sigma$};
    \node [gt] (ibox2) at (-1.5,-0.75) {$\sigma$};
    \node [gt, minimum width=1cm] (ibox3) at (-0.5,-0.75) {Tanh};
    \node [gt] (ibox4) at (0.5,-0.75) {$\sigma$};

    \node [operator] (mux1) at (-2,1.5) {$\times$};
    \node [operator] (add1) at (-0.5,1.5) {+};
    \node [operator] (mux2) at (-0.5,0) {$\times$};
    \node [operator] (mux3) at (1.5,0) {$\times$};
    \node [function] (func1) at (1.5,0.75) {Tanh};

    \node[ct, label={[mylabel]previous cell state}] (c) at (-4.5,1.5) {$c_{i}$};
    \node[ct, label={[mylabel]previous hidden state}] (h) at (-4.5,-1.5) {$h_{i}$};
    \node[ct, fill=cyan!10, label={[mylabel]left:input}] (x) at (-2.5,-3) {$y(t_{i}) $};

    \node[ct, label={[mylabel]new cell state}] (c2) at (4.5,1.5) {$c_{i+1}$};
    \node[ct, label={[mylabel]new hidden state}] (h2) at (4.5,-1.5) {$h_{i+1}$};
    \node[ct, label={[mylabel]left:output}] (x2) at (2.5,3) {$h_{i+1}$};
    \node[ct, fill=yellow!20, label={[mylabel]left:prediction}] (x_t1) at (2.5,4) {$\begin{bmatrix}\Tilde{y}(t_{i+1})\\ \Tilde{\xi}(t_{i+1}) \end{bmatrix}$};
    \draw [ArrowC1] (c) -- (mux1) -- (add1) -- (c2);

    \draw [ArrowC2] (h) -| (ibox4);
    \draw [ArrowC1] (h -| ibox1)++(-0.5,0) -| (ibox1); 
    \draw [ArrowC1] (h -| ibox2)++(-0.5,0) -| (ibox2);
    \draw [ArrowC1] (h -| ibox3)++(-0.5,0) -| (ibox3);
    \draw [ArrowC1] (x) -- (x |- h)-| (ibox3);

    \draw [->, ArrowC2] (ibox1) -- (mux1);
    \draw [->, ArrowC2] (ibox2) |- (mux2);
    \draw [->, ArrowC2] (ibox3) -- (mux2);
    \draw [->, ArrowC2] (ibox4) |- (mux3);
    \draw [->, ArrowC2] (mux2) -- (add1);
    \draw [->, ArrowC1] (add1 -| func1)++(-0.5,0) -| (func1);
    \draw [->, ArrowC2] (func1) -- (mux3);

    \draw [-, ArrowC2] (mux3) |- (h2);
    \draw (c2 -| x2) ++(0,-0.1) coordinate (i1);
    \draw [-, ArrowC2] (h2 -| x2)++(-0.5,0) -| (i1);
    \draw [-, ArrowC2] (i1)++(0,0.2) -- (x2);

\end{tikzpicture}
}
\caption{Schematic representation of the LSTM cell structure}
\label{fig:lstm_cell}
\end{figure}
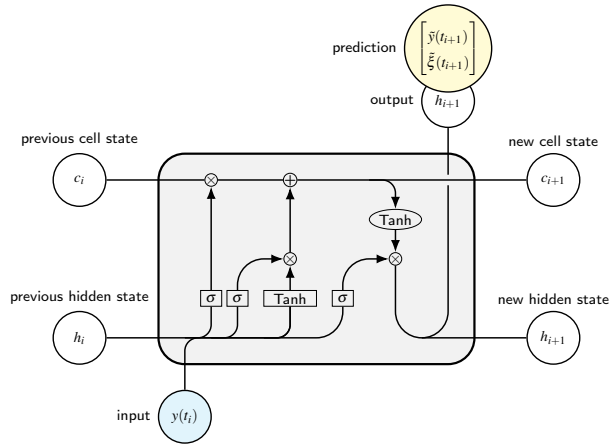

The long short-term memory networks are characterized by a cell state $\bm{c}_{i} \in \mathbb{R}^{N_{h}}$ and a hidden state $\bm{h}_{i} \in \mathbb{R}^{N_{h}}$ that are updated at each step. 
In the case of partial observations, the states are updated by using the observed variables $\bm{y}(t_i)$ as 
  \begin{align}\label{eq:LSTM_states}
    \bm{i}_{i+1} &= \sigma \left(\bm{W}^i [\bm{y}(t_i); \bm{h}_{i}] + \bm{b}^i \right), \\
    \bm{f}_{i+1} &= \sigma \left(\bm{W}^f [\bm{y}(t_i); \bm{h}_{i}] + \bm{b}^f \right), \\
    \bm{o}_{i+1} &= \sigma \left(\bm{W}^o [\bm{y}(t_i); \bm{h}_{i}] + \bm{b}^o \right),\\
    \bm{\Tilde{c}}_{i+1} &= \tanh{\left(\bm{W}^{\bm{g}} [\bm{y}(t_i); \bm{h}_{i}] + \bm{b}^{\bm{g}} \right) }, \nonumber\\ 
    \bm{c}_{i+1} &= \bm{f}_{i+1}*\bm{c}_{i} + \bm{i}_{i+1}*\bm{\Tilde{c}}_{i+1}, \\
    \bm{h}_{i+1} &= \tanh{\left(\bm{c}_{i+1} \right)} * \bm{o}_{i+1}, \nonumber
  \end{align}
where $*$ denotes the elementwise multiplication, \textcolor{black}{and $\sigma$ refers to the sigmoid activation function\cite{Hochreiter_1997_LongShortTermMemory}.} In the LSTM,  $\bm{i}_{i+1}, \bm{f}_{i+1}, \bm{o}_{i+1} \in \mathbb{R}^{N_{h}}$ are the input, forget and output gates (Fig.~\ref{fig:lstm_cell}). The matrices $\bm{W}^i,$ $ \bm{W}^f,$ $ \bm{W}^o,$ $ \bm{W}^{\bm{g}} \in \mathbb{R}^{N_{h} \times (D +N_h)}$ are the corresponding weight matrices, and $\bm{b}^i,$ $\bm{b}^f,$ $\bm{b}^o,$ $\bm{b}^{\bm{g}} \in \mathbb{R}^{N_{h}}$ are the biases. 
The full prediction on the next time step, $\bm{\Hat{x}}(t_{i+1})= [\bm{\Hat{y}}(t_{i+1}), \bm{\Hat{\xi}}(t_{i+1}) ]$ is obtained by applying a dense layer to the hidden state $\bm{h}_{i+1}$, i.e. 
\begin{align*}
    \begin{bmatrix} \bm{\Hat{y}}(t_{i+1}) \\ \bm{\Hat{\xi}}(t_{i+1})  \end{bmatrix}= \bm{W}^{dense} \bm{h}_{i+1} + \bm{b}^{dense},
\end{align*}
where $\bm{W}^{dense} \in \mathbb{R}^{D\times N_h}$ and $\bm{b}^{dense} \in \mathbb{R}^{D} $.
\begin{figure}[h]
    \centering
    \scalebox{0.5}{
  \begin{tikzpicture}[x=3cm, y=1.5cm, >=Stealth]
    \foreach \jlabel [count=\j, evaluate={\k=int(mod(\j-1,1)); \jj=int(\j-1);}]
      in {0, 1}{
        \foreach \ilabel [count=\i] in {1}
            \node [neuron, fill=gray!10, align=left] at (\j, 1-\i) (h-\i-\j){ $\mathbf{LSTM}$};   
          \node [fit=(h-1-\j) (h-1-\j), inner sep=0, draw] (b-\j){ } ;
          \node [input_io, below=of b-\j] (v-\j) {$ \bm{y}(t_{ \jlabel})$};
        \node [dense, above=of h-1-\j, align=left] (d-\j) {{{Dense}}};
          \draw [->] (v-\j) -- (b-\j);
          \draw [->] (b-\j.north) -- (d-\j) node [midway, right] {$\bm{h}_{\j}$};
          }
        \node [output_io, above=of d-1] (output1){$   \begin{bmatrix}
         \bm{\Tilde{y}}(t_{1})\\ \bm{\Tilde{\xi}}(t_{1})\end{bmatrix}
           $ };
         \node [output_io, above=of d-2] (output2){$   \begin{bmatrix}
         \bm{\Tilde{y}}(t_{2})\\ \bm{\Tilde{\xi}}(t_{2})\end{bmatrix}
           $ };
        \node [right=1.5cm of h-1-2](dots1) {\ldots};
        \node [neuron, fill=gray!10,  align=left] at (4, 0) (h-1-4){$\mathbf{LSTM}$ 
            };
        \node [fit=(h-1-4) (h-1-4), inner sep=0, draw] (b-4){} ;
        \node [input_io, below=of b-4] (v-4) {$  \bm{y}(t_{ n-1})  $};
         \node [right=1.5cm of v-2](dots2) {\ldots};
        \draw [->] (v-4) -- (b-4);
        \draw [->] (h-1-1.east) -- (h-1-2.west)node [midway, above] {$\bm{c}_1, \bm{h}_{1}$} ;
        \draw [->] (h-1-2.east) -- (dots1.west)node [midway, above] {$\bm{c}_2, \bm{h}_{2}$};
        \draw [->] (dots1.east) -- (h-1-4.west)node [midway, above] {$\bm{c}_{n-1}, \bm{h}_{n-1}$};
        \node [dense, above=of b-4, align=left] (d-4) {{{Dense}} };
        \draw [->] (b-4.north) -- (d-4.south)node [midway, right] {$\bm{h}_{n}$} ;
        \node [output_io,  above=of d-4] (output){$   \begin{bmatrix}
         \bm{\Tilde{y}}(t_{ n})\\ \bm{\Tilde{\xi}}(t_{ n})\end{bmatrix}
       $ };
        \draw [->] (d-4.north) -- (output.south);
        \draw [->] (d-1.north) -- (output1.south);
        \draw [->] (d-2.north) -- (output2.south);
     \draw[|->] (0.2,0) -- (b-1.west)node [midway, above] {$\bm{c}_0, \bm{h}_{0}$};
    \end{tikzpicture}}
    \caption{LSTM in open-loop configuration: Each cell receives input from the training data.}
    \label{fig:open_loop}
    \end{figure}
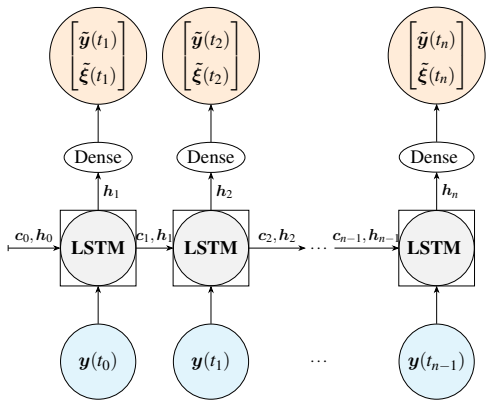

LSTMs are universal approximators for a continuous target function\cite{ hornik1989multilayer, schafer2007recurrent}; however, practically, the network's performance depends on the parameters, such as weights and biases. To determine these parameters, the available data is divided into three subsets: training, validation, and testing. The training and validation data is split into batches of fixed-length time windows. The training data is used for backpropagation through time \cite{werbos1990backpropagation}, and during the forward pass, the network output is computed and compared to a training label using a loss function. In the backward pass, the network parameters are optimized by computing the gradient of the loss function with respect to the parameters with a gradient update via the Adam optimizer \cite{adam}. 
The validation data is employed to optimize hyperparameters and determine parameters before training (e.g. the dimension of the hidden and cell state $N^{h}$), while test data is used to evaluate the final performance of the model.

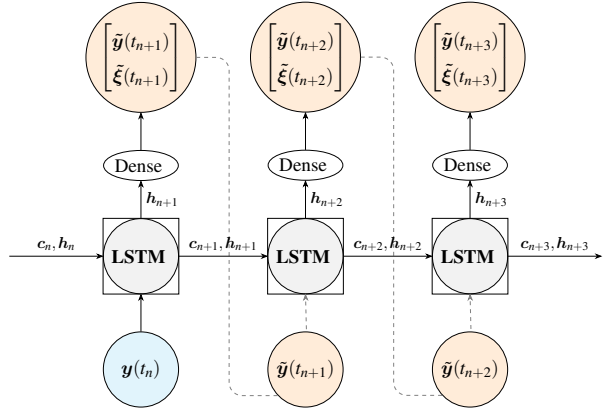
\begin{figure}[h]
    \scalebox{0.5}{
    \begin{tikzpicture}[x=3.5cm, y=1.5cm, >=Stealth]

        \node [neuron, fill=gray!10, align=left] at (4.25, 0) (h-1-4){$\mathbf{LSTM}$ 
            };
        \node [fit=(h-1-4) (h-1-4), inner sep=0, draw] (b-4){} ;
        \node [input_io, below=of h-1-4] (v-4) {$  \bm{y}(t_{ n})  $};
        \draw [->] (v-4) -- (h-1-4);
        \draw [->] (3.25, 0) -- (h-1-4.west)node [midway, above] {$\bm{c}_{n}, \bm{h}_{n}$};
        \node [dense, above=of h-1-4, align=left] (d-4) {{{Dense}} };
        \draw [->] (h-1-4.north) -- (d-4.south)node [midway, right] {$\bm{h}_{n+1}$} ;
        \node [output_io, above=of d-4] (output){$  \begin{bmatrix}
         \bm{\Tilde{y}}(t_{ n+1})\\ \bm{\Tilde{\xi}}(t_{ n+1})\end{bmatrix}
       $ };
        \draw [->] (d-4.north) -- (output.south);
        \node [neuron, fill=gray!10, align=left] at (5.5, 0) (h-1-6){$\mathbf{LSTM}$ 
            };
        \node [fit=(h-1-6) (h-1-6), inner sep=0, draw] (b-6){} ;
        
         \draw[->, rounded corners=10pt, gray, dashed] (output.east) -- ++(0.25, 0) -- ++(0, -6) -- ++(0.6, 0) -- (h-1-6.south);
        \node [input_io, fill=orange!15, below=of h-1-6] (v-6) {$  \Tilde{\bm{y}}(t_{ n+1})  $};
    \draw [->] (h-1-4.east) -- (h-1-6.west)node [midway, above] {$\bm{c}_{n+1}, \bm{h}_{n+1}$};
        \node [dense, above=of b-6, align=left] (d-6) {{{Dense}} };
        \draw [->] (b-6.north) -- (d-6.south)node [midway, right] {$\bm{h}_{n+2}$} ;
        \node [output_io, above=of d-6] (output_n){$ \begin{bmatrix}
         \bm{\Tilde{y}}(t_{ n+2})\\ \bm{\Tilde{\xi}}(t_{ n+2})\end{bmatrix}$ };
       \draw [->] (d-6.north) -- (output_n.south);
        \node [neuron, fill=gray!10, align=left] at (6.75, 0) (h-1-7){$\mathbf{LSTM}$ 
            };
        \node [fit=(h-1-7) (h-1-7), inner sep=0, draw] (b-7){} ;        
         \draw[->, rounded corners=10pt, gray, dashed] (output_n.east) -- ++(0.25, 0) -- ++(0, -6) -- ++(0.6, 0) -- (h-1-7.south);
        \node [input_io, fill=orange!15, below=of h-1-7] (v-7) {$  \Tilde{\bm{y}}(t_{ n+2})  $};
    \draw [->] (h-1-6.east) -- (h-1-7.west)node [midway, above] {$\bm{c}_{n+2}, \bm{h}_{n+2}$};
        \node [dense, above=of b-7, align=left] (d-7) {{{Dense}} };
        \draw [->] (b-7.north) -- (d-7.south)node [midway, right] {$\bm{h}_{n+3}$} ;
        \node [output_io, above=of d-7] (output_n1){$  \begin{bmatrix}
         \bm{\Tilde{y}}(t_{n+3})\\ \bm{\Tilde{\xi}}(t_{n+3})\end{bmatrix}$ };
       \draw [->] (d-7.north) -- (output_n1.south);
       
    \draw [->] (h-1-7.east) -- (7.75, 0)node [midway, above] {$\bm{c}_{n+3}, \bm{h}_{n+3}$};
            
    \end{tikzpicture}}
    \caption{LSTM in closed-loop configuration: The network prediction is used as input for the next cell.}
    \label{fig:closed_loop}
\end{figure}

The network works in an open-loop configuration during training and validation (Fig.~\ref{fig:open_loop}) with the output at each time step depending on current and previous inputs within the time window, and the LSTM's states are reset \textcolor{black}{to zero} at the start of each input sequence. This is also known as "teacher-forced learning" \cite{gers2002learning}, which facilitates a straightforward formulation of loss through backpropagation, as elaborated in Section \ref{sec:loss_formulation}. After training, the network is evaluated on test data with fixed weights and biases, operating in a closed-loop configuration (Fig.~\ref{fig:closed_loop}). \textcolor{black}{After a warm-up of one time window, the network can make long-term predictions in closed loop mode,} even if no data is available, and the states of the LSTM are retained across time steps. The network predicts both the observed and unobserved (hidden) variables. Subsequently, the observed variables are provided as inputs for the next time step, which allows the autonomous evolution of the LSTM.
\section{Tasks and loss functions}\label{sec:loss_formulation}
The LSTM's internal dynamics, and therefore its prediction, are determined by the weights and biases of the model, which are updated during the training. The loss function plays a critical role in this process by defining the optimization problem that guides the updating of these parameters. The choice of the loss function depends on the specific application and can greatly impact the accuracy and generalization performance of the model. In the following, we propose two loss functions to guide the LSTM training: (i) a data-driven loss, and (ii) a physics-informed loss. 

\subsection{Low-to-high resolution loss (LH-LSTM)}
The first task is to infer the high-resolution dynamics from low-resolution inputs, thereby reducing the cost of acquiring full measurement data. In fluid dynamics, in which problems display turbulent behaviour, this task becomes especially difficult as the prediction is sensitive due to the chaotic nature of the data. The network predicts the full state $\bm{\bm{\Hat{x}}}(t_{i}) = [\bm{\bm{\Hat{y}}}(t_{i}); \bm{\Hat{\xi}}(t_{i})]$ given a partial input $\bm{y}(t_{i})$. 
In this section, we assume that the full label $\bm{x}(t_{i}) =  [\bm{y}(t_{i}); \bm{\xi}(t_{i})]$ is only available for a short amount of time, which can be employed for training and validation, and partial data is available for the remaining time. Hence, we can define a data-driven loss using a mean-squared error (MSE) 
\begin{align}
    \mathcal{L}_{dd}(\bm{x}, \bm{\Hat{x}}) = \frac{1}{N_t} \sum_{i=1}^{N_t} \|\bm{x}(t_{i}) -\bm{\Hat{x}}(t_{i}) \|^2,
    \label{eq:DDLoss_fullstate}
\end{align} 
where $||\cdot||$ is the $L^2$-norm. 
Although simple to implement, data-driven loss functions have limitations. 
Specifically, they are sensitive to overfitting, especially when the training data is noisy. \textcolor{black}{To mitigate overfitting, we employ Tikhonov regularization by adding 
 $\alpha_{\textit{tikh}}\|\bm{\Hat{x}}\|^2, \alpha_{\textit{tikh}}\in \mathbb{R}^{+}$ to Eq.~\eqref{eq:DDLoss_fullstate} \cite{tikhonov1995numerical}.} When data is limited, these loss functions may struggle to generalize to regions of the attractor that are sparsely sampled during training. To regularize the problem, the incorporation of the physical knowledge or constraints into the loss function is an effective means~\cite{doan2021short,ozalp2023physics}, which favour physical solutions (in contrast to traditional Tikhonov regularization).

\subsection{Physics-informed loss (PI-LSTM)}\label{sec:pi_lstm}

Physics-informed losses have shown success in feed-forward neural networks, in which automatic differentiation can be exploited to accurately compute the derivative with respect to time and space to machine precision \cite{lagaris1998pinn, raissi2019physics}. Unlike feed-forward neural networks, RNNs have a dynamic temporal structure that makes it difficult to accurately differentiate the prediction with respect to time because of the recurrence and the type of inputs. As a result, the governing equations have to be incorporated in a manner that accounts for the temporal structure of the RNN architecture. This approximation is typically achieved by discretizing the time derivative in the loss function using finite differences, which allows the physics-informed loss to be computed but it relies on the numerical differentiation scheme \cite{ozalp2023physics}. 
Hence, the autonomous dynamical system defined by Eq.~\eqref{autonom_dyn_system} is reformulated with its formal solution through the integral 
\begin{align}\label{integral_equation_dynamical_system}
\bm{x}(t_{i+1}) &= \int_{t_0}^{t_{i+1}} f(\bm{x}(t)) dt \
= \bm{x}(t_i) + \int_{t_i}^{t_{i+1}} f(\bm{x}(t)) dt, \ i\geq0,
\end{align}
which enables the use of numerical quadrature methods to approximate the integral $\int_{t_i}^{t_{i+1}} f(\bm{x}(t)) dt$, instead of approximating the derivative $\frac{d\bm{x}(t)}{dt}$ to construct an accurate surrogate model. This approach is naturally compatible with explicit numerical schemes of different orders of accuracy, including the explicit Runge-Kutta family or pseudo-spectral methods, thereby providing a  flexible framework for modelling physical constraints into RNNs.
Based on Eq.~\eqref{integral_equation_dynamical_system}, we define the residual of the dynamical system
\begin{align}
    \mathcal{R}(\bm{x}(t_{i+1})) = \bm{x}(t_{i+1}) - \left( \bm{x}(t_i) + \int_{t_i}^{t_{i+1}} f(\bm{x}(t)) dt\right).
\end{align}
The true solution, $\bm{x}(t)$, of the dynamical system, is such that $\mathcal{R}(\bm{x}(t_i)) = 0$, for all $t_i>t_0$. In this manner, the predictions $\bm{\Hat{x}}$ of the network favour physical solutions by minimizing the physics-informed loss 
\begin{align}
    \mathcal{L}_{pi}(\bm{\Hat{x}}) = \frac{1}{N_t} \sum_{i=0}^{N_t -1} \|   \mathcal{R}(\Hat{\bm{x}}(t_{i+1})) \|^2.
\end{align}

The proposed loss function addresses a distinct task from the data-driven loss \eqref{eq:DDLoss_fullstate} by not requiring a full label, $\bm{x}(t_{i})$, making it particularly advantageous in situations in which full labels are not available. Instead, it enables the network's prediction to be constrained based on the governing equations, which allows the reconstruction and the forecasting of unobserved (hidden) variables. By providing a regularization during the training of the LSTM, we also ensure that the predictions satisfy the governing equation defined in \eqref{autonom_dyn_system} within a numerical tolerance. Further, this regularization helps to constrain the parameter space of the weights, which can mitigate overfitting and improve generalization.
We also include a data-driven loss for the available data by computing the MSE between the network prediction on the partial input $\bm{\Hat{y}}(t_{i+1})$ and the partial label $\bm{y}(t_{i+1})$
\begin{align}\label{dd_loss}
    \mathcal{L}_{dd}(\bm{y}, \bm{\Hat{y}})= \frac{1}{N_t} \sum_{i=1}^{N_t} \| \bm{y}(t_{i}) - \bm{\Hat{y}}(t_{i}) \|^2.
\end{align}
The loss of the physics-informed LSTM is computed by combining the data-driven loss and weighing the physics-informed loss as 
\begin{equation}
    \mathcal{L}(\bm{y}, \bm{\Hat{x}}) = \mathcal{L}_{dd}(\bm{y}, \bm{\Hat{y}})  + \alpha_{pi}\mathcal{L}_{pi}(\bm{\Hat{x}}), \quad \alpha_{pi} \in \mathbb{R}^{+}, 
    \label{eq:totPILoss}
\end{equation}
where $\alpha_{pi}$ is a penalty hyperparameter, which acts as a regularization factor. \textcolor{black}{In the systems under investigation, there is no obvious a-priori choice for the selection of an appropriate regularization hyperparameter. Poor selection of $\alpha_{pi}$ can result in suboptimal models. If $\alpha_{pi}$ is too large, it may lead to overfitting of the physics whilst disregarding the observations. Conversely, if $\alpha_{pi}$ is too small, the models may overfit the measurements and fail to generalize. To select the regularization factor, $\alpha_{pi}$, we employ a grid search as detailed in the Appendix~\ref{sec:appendix}.} 

\section{Inferring the stability of the dynamical system from data}\label{sec:stab_prop}

Thanks to its recurrent nature, the LSTM can use its own prediction as input in the closed-loop mode, which allows it to predict the system's full state in time beyond the input. This autonomous evolution of LSTMs defines a dynamical system. 
 Mathematically, we assess the LSTM stability properties by studying their tangent space \cite{Margazoglou2023}. By introducing small perturbations to the system's trajectory and linearizing the system's equations, we can compute its stability in different directions. We determine the properties of its linear tangent space, such as Lyapunov exponents (LEs) and covariant Lyapunov vectors (CLVs). To gain insight into the network's dynamical behaviour, we apply these concepts to the LSTM by mathematically deriving the LSTM Jacobian.

\subsection{Lyapunov exponents}
The solution of the dynamical system in Eq.~\eqref{autonom_dyn_system}
 is $\bm{x}(t)$, which defines a trajectory in the phase space of dimension $D$.
To analyse the properties of the dynamical system and its attractor, we impose infinitesimal perturbations to the trajectory of the system as
\begin{equation}\label{first_order_pertubation}
    \bm{x} + \bm{u},\quad \bm{x} \sim \mathcal{O}(1),\quad \bm{u} \sim \mathcal{O}(\epsilon),\quad \epsilon \to 0.
\end{equation}
By substituting \eqref{first_order_pertubation} into \eqref{autonom_dyn_system} and linearizing the system, the perturbation is governed by the tangent equation
\begin{equation}\label{tangent_equation}
    \frac{d\bm{u}(t)}{dt} = \bm{J}(\bm{x}(t))\bm{u}(t),
\end{equation}
where $J_{ij} = \frac{\partial f_i(\bm{x})}{\partial x_j}$ are the components of the Jacobian $\bm{J}(\bm{x}(t)) \in \mathbb{R}^{D\times D}$, which is time-dependent on chaotic attractors. This shows that the perturbation $\bm{u}(t)$ evolves along the tangent space with respect to $\bm{x}(t)$, and must therefore be integrated alongside Eq.~\eqref{autonom_dyn_system}.
Considering $K$ random perturbations, we numerically integrate $K\leq D$ tangent vectors $\bm{u}_i \in \mathbb{R}^D$ as columns of $\bm{U} \in \mathbb{R}^{D \times K}$ 
\begin{equation}\label{tangent_equation_matrix}
\frac{d\bm{U}}{dt} = \bm{J}(\bm{x}(t))\bm{U}.
\end{equation}
In chaotic systems, the tangent vectors align exponentially fast with the leading Lyapunov vector, which can lead to an ill-conditioned matrix $\bm{U}$~\cite{Ginelli2013_clv, Sandri_1996_NumericalCalculationLyapunovExponents}. For practical purposes, $\bm{U}$ is periodically orthonormalized through the Gram-Schmidt procedure \cite{benettin1980lyapunov_p2}. Hence, we decompose $\bm{U}$ into an orthogonal matrix $\bm{Q}$ and an upper-triangular matrix $\bm{R}$ with QR-decomposition, i.e. $\bm{U}(t)= \bm{Q}(t)\bm{R}(t)$. The columns of $\bm{Q}$ form an orthonormal basis for the tangent space and are known as Gram-Schmidt vectors (GSVs). The diagonal entries $R_{ii}(t)$ correspond to the local growth rates of the corresponding GSV $\bm{q}_i(t)$.
The Lyapunov exponents (LEs) are computed by taking the time average of the logarithms of $R_{ii}(t)$ 
\begin{equation}\label{lambda_i}
    \lambda_i = \lim_{T \to \infty}\frac{1}{T-t_0}  \int_{t_0}^{T} \ln\left( R_{ii}(t)\right).
\end{equation}
The Lyapunov spectrum, $\lambda_1 \geq \dots \geq \lambda_D$, provides fundamental insight into the chaotic properties and geometry of an attractor~\cite{Ginelli2013_clv,huhn2020stability}. If the leading Lyapunov exponent $\lambda_1<0$, the perturbation decay and the attractor is a fixed point. If $\lambda_1 = 0$ and the remaining exponents are negative, the attractor is a periodic orbit. If $\lambda_1> 0 $, the perturbation grows exponentially and, typically, the attractor is chaotic. In this case, the Lyapunov time $\tau_{\lambda} = \frac{1}{\lambda_1}$ defines a characteristic timescale for two nearby trajectories to separate, which gives an estimate of the system’s predictability horizon~\cite{boffetta2002predictability,Nastac_2017_LE_Metric}. Further, the LEs provide an estimate for the attractor dimension through the Kaplan-Yorke dimension~\cite{frederickson1983liapunov_conj, kantz_schreiber_2003}
\begin{equation}
    D_{KY} = k + \frac{\sum_{i=1}^k \lambda_i}{|\lambda_{k+1}|}
\end{equation}
 with $\sum_{i=1}^k \lambda_i > 0 $ and $\sum_{i=1}^{k+1} \lambda_i < 0 $.
 
 \subsection{Covariant Lyapunov vectors}
 \label{sec:CLVs}
Because LEs are scalars, 
 they cannot describe the geometry of the tangent space. For this, we need to find a suitable basis that spans the tangent space.
 The GSVs $\bm{Q}(t)$ provide an orthogonal basis of the tangent space, which is not time-reversible because of the frequent orthogonalizations. On the other hand, there exists a coordinate-independent, local decomposition of the phase space into subspaces spanned by the covariant Lyapunov vectors (CLVs)~\cite{Oseledec_1968_MultiplicativeErgodicTheorem, ruelle1979_ergodictheory}. Unlike the GSVs, which are orthonormal by construction, the subspaces spanned by the CLVs are generally non-orthogonal to each other, which provide information on the local geometric structure of the chaotic attractor. If the CLVs $\bm{V}= \begin{bmatrix} \bm{v}_1, \bm{v}_2, \dots, \bm{v}_k\end{bmatrix}$ are uniquely defined (i.e.~non degenerate), then each CLV $\bm{v}_i\in \mathbb{R}^D$ describes the individual expansion and contraction associated with the LE $\lambda_i$. The CLVs can be recovered from the local growth rates $\bm{R}(t)$ and the GSVs $\bm{Q}(t)$ by evolving backwards in time after the forward-in-time simulation is completed. We provide a brief overview of the computation of the CLVs in the Appendix~\ref{subsec:appendix_clvs}, for further details on the computation, we refer the reader to \cite{Ginelli2007_CharacterizingDynCLVs,Ginelli2013_clv,Margazoglou2023}.

Two key pieces of information are contained in the CLVs.
First, CLVs provide information on the hyperbolicity of the chaotic attractor \cite{Eckmann_Ruelle_1985}. 
An attractor is hyperbolic if there is a splitting of tangent space at every point of the trajectory $\bm{x}(t)$ into three subspaces: unstable $E^U_{\bm{x}}$, neutral $E^N_{\bm{x}}$, and stable $E^S_{\bm{x}}$. These subspaces contain CLVs associated with positive, zero, and negative LEs, respectively. This splitting has profound implications for the geometric structure of the attractor and in hyperbolic systems when there are no tangencies between these subspaces~\cite{huhn2020stability}. Therefore, the distribution of the angles between these subspaces is bounded away from zero, which is a property that can be analysed with the CLVs. 
Second, CLVs provide information on the most important directions in phase space, which is useful for reduced-order modelling. 
CLVs can split the tangent space of spatially extended dissipative systems into a decomposition of a fixed, finite number of ``physical'' modes and a remaining set of "spurious" modes \cite{takeuchi2011hyperbolic}. The physical modes contain the relevant dynamics of the phase space and define a finite-dimensional manifold in the tangent space, while the spurious modes correspond to the negative LEs and increase in number with increasing resolution. These spurious modes are hyperbolically decoupled from the physical modes, because perturbations along them decay quickly and do not propagate to the physical modes. A way to characterize the transition between physical and spurious modes is by analyzing the statistics of the CLV angles and locating a clear absence of tangencies in pairs of adjacent CLVs~\cite{takeuchi2011hyperbolic}. The number of physical modes provides a lower bound for the degrees of freedom in the simulation.


\subsection{Jacobian of the LSTM}
We perform stability analysis on the dynamical system defined by the LSTM. Key to the analysis is the Jacobian of the system, therefore, the Jacobian of the LSTM is required. To compute the Jacobian, we express in a compact form the LSTM equations ~\eqref{eq:LSTM_states} during closed-loop as 

\begin{align}\label{eq:LSTM_map}
 [ \bm{c}_{i+1}, \bm{h}_{i+1}]^T &= LSTM(\bm{\Hat{x}}(t_{i}), \bm{c}_{i}, \bm{h}_{i}), \\
     \bm{\Hat{x}}(t_{i+1})  &= \bm{W}^{dense} \bm{h}_{i+1} + \bm{b}^{dense}.
\end{align}
The Jacobian is the gradient of the internal states at a single timestep 
\begin{align}\label{coupled_Jac}
    \bm{J}_{LSTM}( \bm{c}_{i},  \bm{h}_{i}) = \begin{bmatrix} \frac{\partial \bm{c}_{i}}{\partial \bm{c}_{i-1}}  &\frac{\partial \bm{c}_{i}}{\partial \bm{h}_{i-1}} \\[2ex]
    \frac{\partial \bm{h}_{i}}{\partial \bm{c}_{i-1}} & \frac{\partial \bm{h}_{i}}{\partial \bm{h}_{i-1}} \end{bmatrix}, 
\end{align}
which analytically is
\begin{align}
\begin{split}
    \frac{\partial \bm{c}_{i}}{\partial \bm{c}_{i-1}} &= \mathbf{I}  *\bm{f}_{i},\\
\frac{\partial \bm{c}_{i}}{\partial \bm{h}_{i-1}} &=
    \bm{c}_{i} *\bm{f}_{i}* ( \mathbf{I}-\bm{f}_{i}) \bm{W}^f + \bm{i}_{i} *( \mathbf{I}-\bm{i}_{i}) \bm{W}^i *\Tilde{\bm{c}}_i +\bm{i}_i * \left(\mathbf{I}-\bm{\Tilde{C}}_{i}^2\right) \bm{W}^{\bm{g}},\\
\frac{\partial \bm{h}_{i}}{\partial \bm{c}_{i-1}} &= \left( \mathbf{I} -            \tanh^2{\left(\bm{c}_{i} \right)} \right)*\bm{o}_{i} * f_i,\\
\frac{\partial \bm{h}_{i}}{\partial \bm{h}_{i-1}} &=  
    \bm{o}_{i}*( \mathbf{I}-\bm{o}_{i})*\tanh{(\bm{c}_{i})} + 
    \left(  \mathbf{I} - \tanh^2{(\bm{c}_{i})}\right)*\bm{o}_{i} * \\
    &\left(\bm{c}_{i} *\bm{f}_{i} *( \mathbf{I}-\bm{f}_{i}) \bm{W}^f + \bm{i}_{i} *( \mathbf{I}-\bm{i}_{i}) \bm{W}^i *\Tilde{\bm{c}}_i +\bm{i}_i * \left( \mathbf{I}-\bm{\Tilde{C}}_{i}^2\right) \bm{W}^{\bm{g}}\right).
    \end{split}
    \label{eq:LSTM_Jac_analytical}
\end{align}

where $\mathbf{I}\in \mathbb{R}^{N_{h}\times N_{h}}$ is the identity matrix and $\bm{J}_{LSTM} \in \mathbb{R}^{(N_{h}+N_{h})\times (N_{h}+N_{h})}$. In chaotic systems, the Jacobian matrix varies at each time step. To analyse the stability from partial observations, the idea is to analyse the LSTM Jacobian for the computation of the LEs and CLVs. With this, we can compute the stability properties in the network in the closed-loop mode. From Eq.~\eqref{eq:LSTM_Jac_analytical} we expect a maximum of $2N_{h}>D$ Lyapunov exponents, of which the first $D$ are physically relevant, whereas the remaining are spurious (i.e., they depend on the network's architecture). We provide a pseudocode for the calculation \textcolor{black}{in Algorithm~\ref{alg:lstm_le} in the Appendix~\ref{subsec:appendix_comp_le}.}

\section{Results} \label{sec:results}
The objective is to assess the capabilities of the two LSTM architectures under investigation in the forecasting of both observed and unobserved (hidden) chaotic time series. \textcolor{black}{
 To study the forecast performance of the proposed LSTMs, we employ the prediction horizon during the closed-loop configuration
   \begin{equation}\label{pred_horizon}
        \frac{ \|\bm{x}(t_{N_{PH}}) - \bm{\Hat{x}}(t_{N_{PH}}) \|}{\sqrt{\frac{1}{N_{PH}} \sum_{i={0}}^{N_{PH}} \|\bm{x}(t_i)\| ^2}} < k,
    \end{equation}
with $k=0.5$ as in \cite{Vlachas_2020_backprop}. Starting from the same initial conditions, the prediction horizon quantifies how far in the future the LSTM can follow autonomously the reference chaotic solution. }
We then perform a long autonomous temporal evolution of the trained PI-LSTM and LH-LSTM networks, and compare the statistics with the target chaotic dynamical systems. In Sec.~\ref{sec:KS}, we fix the number of unobserved (hidden)  variables in the example of the Kuramoto-Sivashinsky equation. After reconstructing the full state, we compute the statistics and spectrum of the kinetic energy, as well as key stability properties such as the Lyapunov exponents, and the angles between selected CLVs. In Sec.~\ref{sec:l96}, we employ the Lorenz~96 equation and test the capabilities of the LSTM networks when a different number of hidden variables is considered.

\subsection{Kuramoto–Sivashinsky} \label{sec:KS}
The Kuramoto–Sivashinsky equation, also known as the KS equation, was first derived in \cite{Kuramoto_1978_DiffusionInducedChaos} to describe diffusion-induced space-time chaos in reaction systems. In \cite{Sivashinsky_1977_NonlinearAnalysisHydrodynamicInstability}, the equation was independently derived to model small diffusive-thermal instabilities of laminar flame plane fronts. The KS equation is a fourth-order nonlinear partial differential equation that can be written in one spatial dimension as 
\begin{equation}\label{ks_equation}
\bm{\phi}_t(t,\mathrm{x}) + \bm{\phi}_{\mathrm{x}\mathrm{x}}(t,\mathrm{x}) + \bm{\phi}_{\mathrm{x}\mathrm{x}\mathrm{x}\mathrm{x}}(t,\mathrm{x}) + \bm{\phi}(t,\mathrm{x}) \bm{\phi}_{\mathrm{x}}(t,\mathrm{x}) = 0,
\end{equation}
with $\mathrm{x} \in[0, L)$ being the spatial direction. By assuming periodic boundary conditions $\bm{\phi}(t, 0) = \bm{\phi} (t, L)$ and setting $L=2\pi\cdot10$, the solution of the KS equation displays chaotic behaviour \cite{Nicolaenko_1985_KSE_Stability_Attractors, Hyman_1986_KuramotoSivashinsky} with $\lambda_1 = 0.08$. We spatially discretize Eq. \eqref{ks_equation} with $128$ degrees of freedom, corresponding to $\Delta \mathrm{x} = L/128$, and select a timestep $\Delta t = 0.25$. The equation is then solved with a fourth-order spectral scheme for stiff PDEs \cite{Kassam_2005_fourth_order, Pathak_2017_ml_le} up to $T = 2.5 \cdot 10^4$, resulting in $ 10^5$ samples. To eliminate transients, we disregard the initial $1000$ samples and partition the remaining data into training, validation, and testing sets with sizes $5\cdot10^4$, $2\cdot 10^4$, and $5\cdot10^4$, respectively. The tangent space calculation follows \cite{Pathak_2017_ml_le}, and the LEs and the CLVs are computed, resulting in a Kaplan-Yorke dimension equal to $D_{KY}=15.02$. We analyse the distribution of angles between the CLVs, and detect $j=29$ physical modes for the selected parameters in the present study; hence, the modes $j\geq 30$ are spurious. 
\begin{figure}[t]
    \centering
    \includegraphics{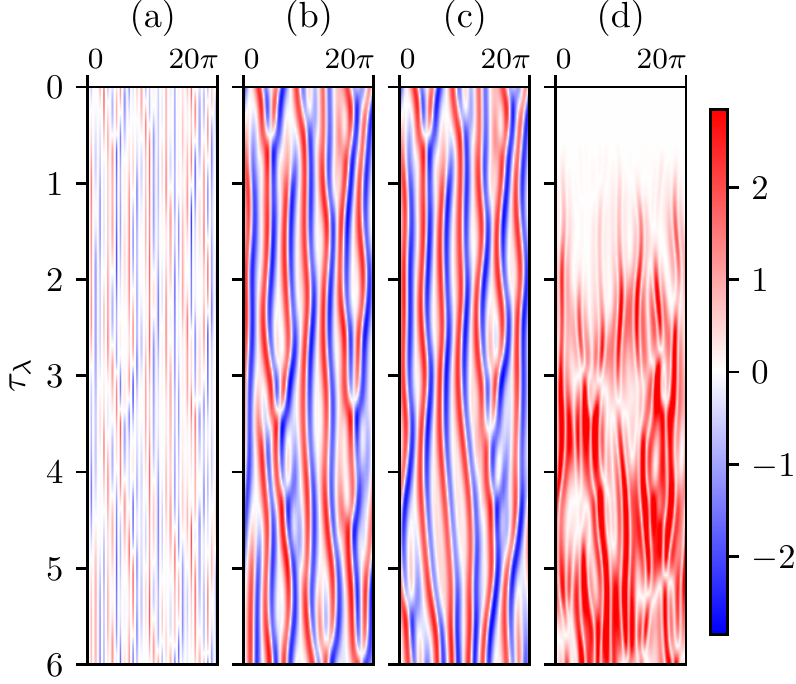}
    \caption[KS: Spatio-temporal forecast]{Comparison of a test data set's (a) input, (b) label/reference solution, (c) LH-LSTM's closed-loop prediction, and (d) the absolute error between the target and LH-LSTM prediction.}
    \label{fig:ks_short_term_prediction_dd}
\end{figure}
\begin{figure}[t]
    \centering
     \includegraphics{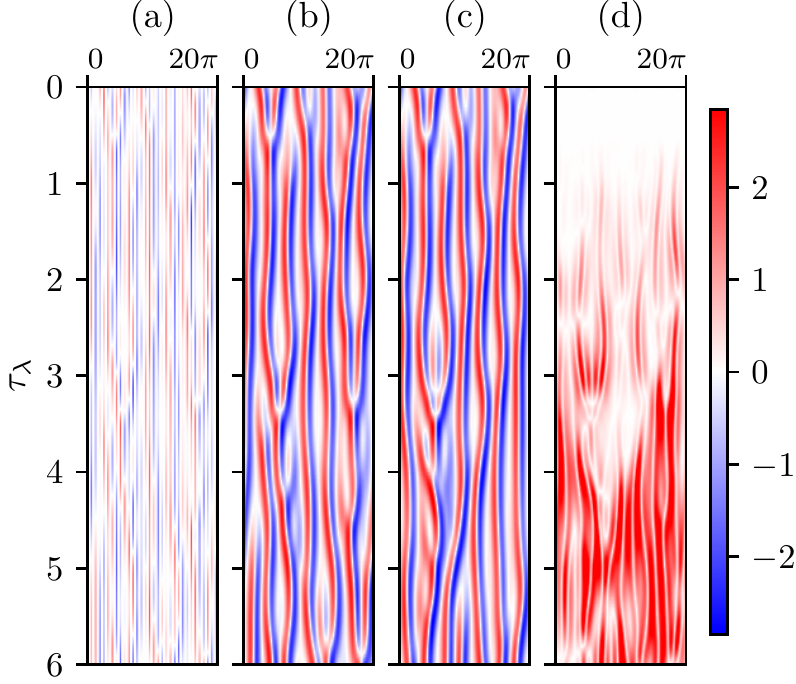}
   \caption[KS: Spatio-temporal forecast]{Comparison of a test data set's (a) input, (b) label/reference solution, (c) PI-LSTM's closed-loop prediction, and (d) the absolute error between the target and PI-LSTM prediction.}
    \label{fig:ks_short_term_prediction_pi}
\end{figure}
In the following, we present the results with $D_y = 32$ observed inputs and the network is trained to predict the full state with $D=128$. \textcolor{black}{We select the window size in $\{10, 20, 50\}$, and the \textcolor{black}{cell and hidden state dimensions $N_h$ } in $\{200, 500 \}$ with a grid search, see Table~\ref{table:ks_hyperparameters} the Appendix~\ref{subsec:appendix_hyperparameters} for details}. We do not observe significant differences in the network performance for different choices of window sizes, and \textcolor{black}{the cell and hidden state dimensions}. For the physics-informed LSTM, the weighing $\alpha_{pi}$ is selected from $\{ 100, 10, 1\}$.
We present the closed-loop prediction of the two LSTMs over $6$\textcolor{black}{$\tau_{\lambda}$}, with the LH-LSTM in Fig.~\ref{fig:ks_short_term_prediction_dd} and the PI-LSTM in Fig.~\ref{fig:ks_short_term_prediction_pi}. Both networks accurately predict the KS solution over a short time horizon, with a prediction horizon of $2.6$\textcolor{black}{$\tau_{\lambda}$} for the data-driven model and $2.86$\textcolor{black}{$\tau_{\lambda}$} for the physics-informed model.

\begin{figure}[ht]
    \centering
    \includegraphics{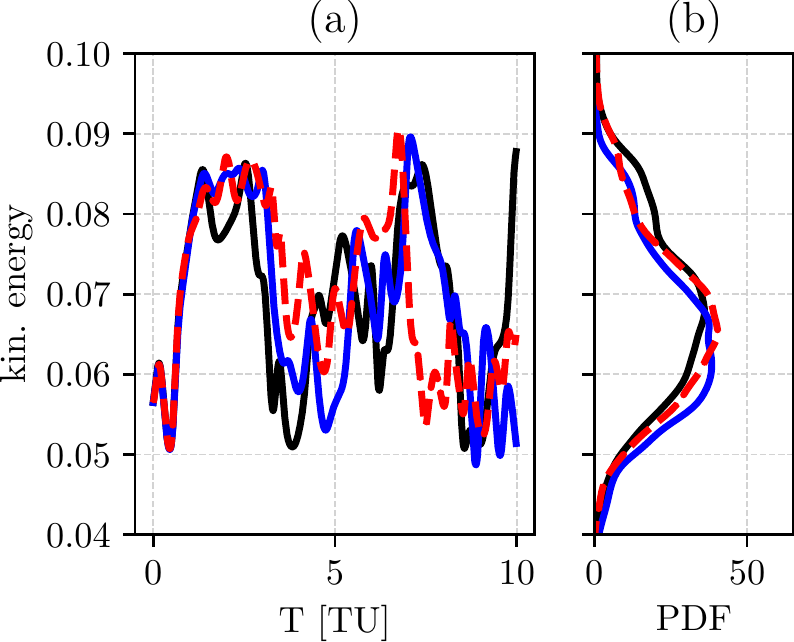}
    \caption[KS: Kinetic energy]{Kinetic energy of the Kuramoto–Sivashinsky system: Comparison of the target (black line), PI-LSTM (red line) and LH-LSTM (blue line) for (a) short time span on the test data and (b) the statistics over $500$\textcolor{black}{$\tau_{\lambda}$}.}
    \label{fig:kin_energy}
\end{figure}

We investigate the long-term physical behaviour of the network prediction by examining the kinetic energy $E(t)=\frac{1}{2L}\int d{\rm x}|\phi(t,{\rm x})|^2$ of the reference solution and the networks' predictions in Fig.~\ref{fig:kin_energy}. Figure~\ref{fig:kin_energy}(a) shows the kinetic energy of the first $10$\textcolor{black}{$\tau_{\lambda}$} of the test set, with an overlap of the energy during the prediction horizon. After the prediction horizon, the kinetic energies gradually separate, due to the chaos, but the LSTMs' \textcolor{black}{closed-loop} predictions have correct statistical behaviour in the kinetic energy, suggesting that networks produce long-term physical solutions (Fig.~\ref{fig:kin_energy}(b)).

\begin{figure}[t]
    \centering
    \includegraphics{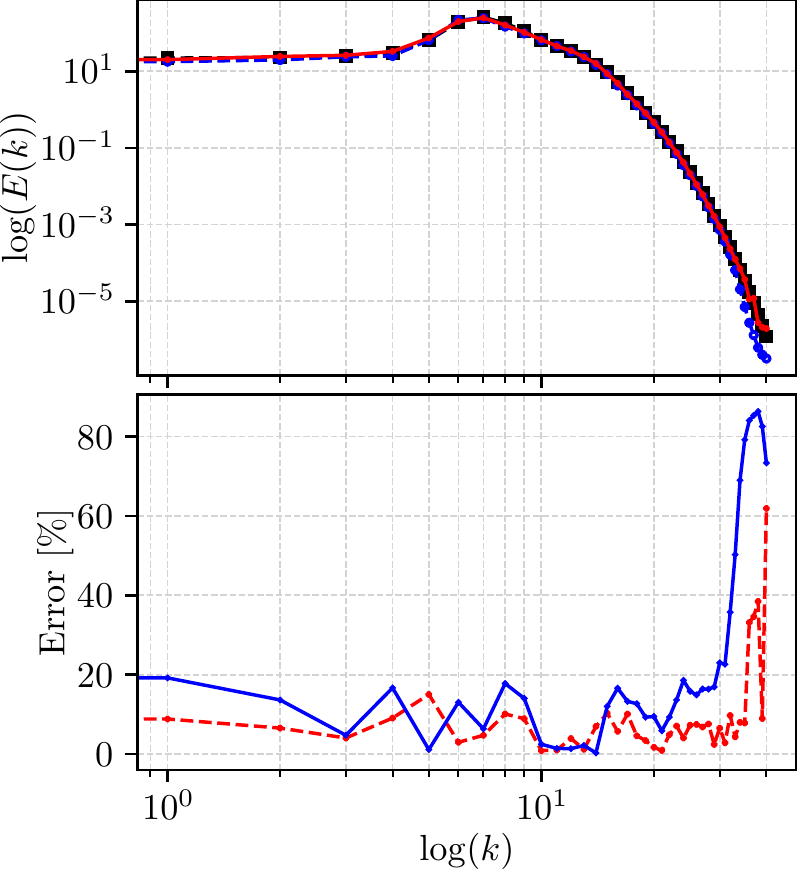}
    \caption[KS: Energy spectrum]{Energy spectrum of the Kuramoto–Sivashinsky system: Comparison of the target (black line), PI-LSTM (red line) and LH-LSTM (blue line). The distribution shows the energy transfer across different scales, with high wavenumbers corresponding to small-scale behaviour. }
    \label{fig:energy_spectrum}
\end{figure}

To investigate this further, we present in Fig.~\ref{fig:energy_spectrum} the time-averaged energy spectrum, defined as $E(k)=\frac{1}{T}\int dt|\phi(t,k)|^2$, which is the distribution of energy at different wavenumbers $k$. High wavenumbers correspond to small-scales, which are difficult to model accurately during the numerical computation, as they involve a large range of spatial and temporal scales. Despite passing only partial spatial observations as an input, both LSTM models accurately capture the energy of the majority of wavenumbers, showing that the models accurately extrapolate to smaller scales. 

\begin{figure}[h]
    \centering
    \includegraphics[width=0.48\textwidth]{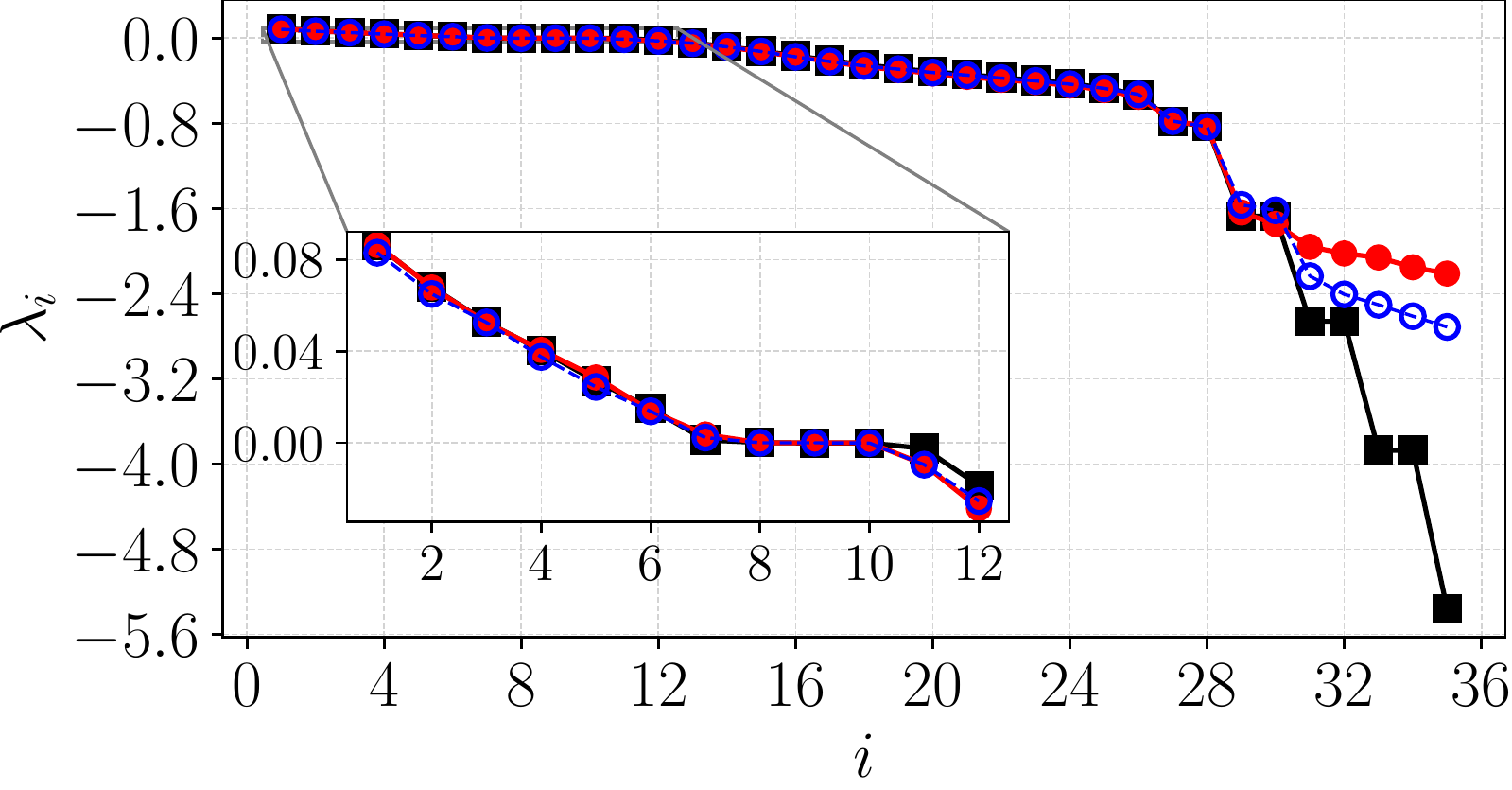}
    \caption[KS: Lyapunov exponents]{The first $35$ Lyapunov exponents of the Kuramoto–Sivashinsky system for the reference data (black squares), PI-LSTM (red dots) and LH-LSTM (blue crosses). Both networks infer correctly the first $30$ Lyapunov exponents, which correspond to the physical behaviour (i.e., physical modes~\cite{takeuchi2011hyperbolic}) of the system.}
    \label{fig:ks_lyapunov_exponents}
\end{figure}

Although the energy spectrum is a global quantity to analyse the distribution of energy across different scales, stability analysis provides a detailed tool to understand the chaotic and tangent dynamics of the system. In Fig.~\ref{fig:ks_lyapunov_exponents}, we compare the reference LEs (in black squares) with the LEs extracted from the LH-LSTM (red circles) and the PI-LSTM (blue crosses), \textcolor{black}{computed with Algorithm~\ref{alg:lstm_le} and hyperparameters in Table~\ref{table:ks_hyperparameters_les}}. (As in\cite{Vlachas_2020_backprop}, the network does not capture two zero LEs ($\lambda_9$ and $\lambda_{10}$) and the plot is augmented, accordingly.) By reconstructing the full state the networks effectively reconstruct the tangent space, the properties of which are encapsulated in the LEs. Both networks infer the first $29$ LEs with high accuracy, with an error of $3.7\%$ (LH-LSTM) and $0.2\%$ (PI-LSTM) in $\lambda_1$. This means that the machine has inferred the correct physical modes based on the approach in \cite{takeuchi2011hyperbolic}. The PI-LSTM case can also infer the hidden variables without fully labelled data, which provides further evidence of the physical accuracy and correct dynamics.

\begin{figure*}[t]
    \centering
    \includegraphics[width=0.6\textwidth]{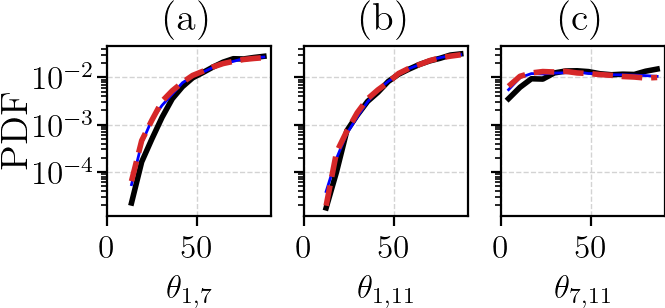}
    \caption[KS: CLV angles]{The angle distribution of the Kuramoto–Sivashinsky system for the leading covariant Lyapunov vectors of three different subspaces: (a) unstable-neutral, (b) unstable-stable, and (c) neutral-stable. The black line corresponds to the reference data, the red and blue lines indicate the results obtained from the \textcolor{black}{$10000 \tau_{\lambda}$ long autonomous evolution of the} PI-LSTM and LH-LSTM models, respectively.}
    \label{fig:ks_clv_angels}
\end{figure*}

We examine the principal angles between the dominant CLVs of the unstable subspace $E_x^U$, the neutral subspace $E_x^N$, and the stable subspace $E_x^S$ in Fig.~\ref{fig:ks_clv_angels}, where 
\begin{equation}
	\theta_{a,b} = \frac{180^{\circ}}{\pi} \cos^{-1}(|\bm{v}_a* \bm{v}_a|),
	\label{eq:clvangles}
\end{equation} 
$\theta_{a,b} \in [0^{\circ},90^{\circ}]$. 
In Fig.~\ref{fig:ks_clv_angels}, the agreement of the angle statistics between reference and the LSTMs is within a negligible numerical error, which reflects the robust and accurate learning of the ergodic properties from the partial input data. Further discussion on the CLVs is provided in \textcolor{black}{the Appendix~\ref{subsec:appendix_clvs_ks} and Fig.~\ref{fig:KS_clv_angles_appendix}.}  
\FloatBarrier
\subsection{Lorenz-96}\label{sec:l96}
\begin{figure}[t]
    \centering
     \includegraphics[width=0.5\textwidth]{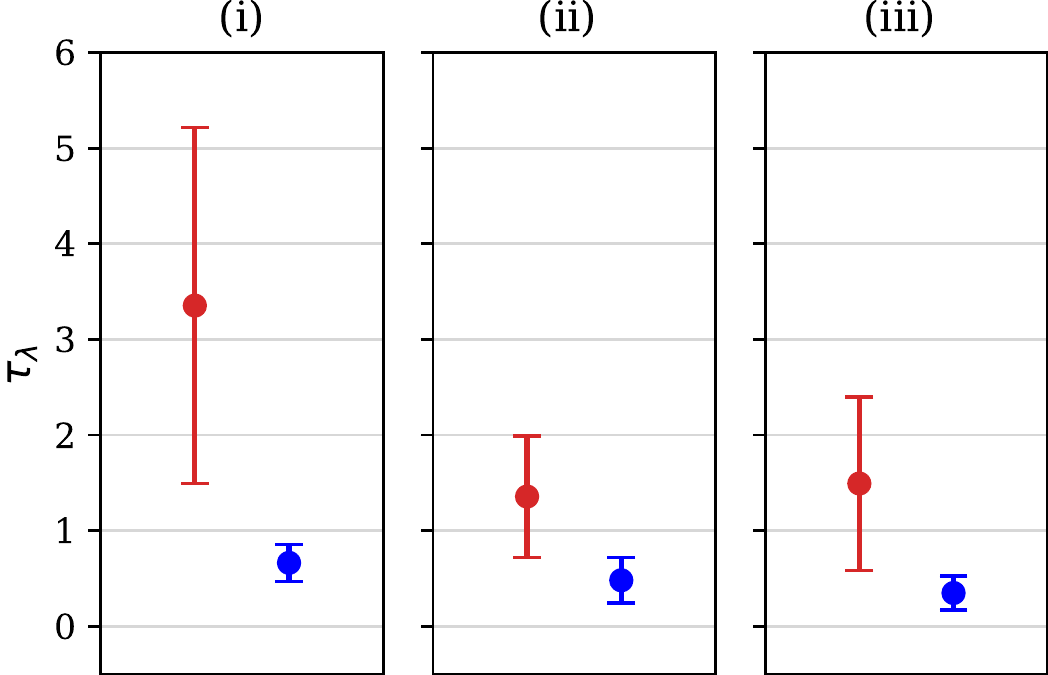}
   \caption[Lorenz-96: Prediction Horizon]{\textcolor{black}{Mean and standard deviation of the prediction horizon for PI-LSTM (red line) and LH-LSTM (blue line) for $100$ predictions in the test set with (i) $D_{\xi} = 2$, (ii) $D_{\xi} = 6$, (iii) $D_{\xi} = 10$ missing variables in the closed-loop configuration.}}
    \label{fig:l96_testcase_predictionhorizon}
\end{figure}

The Lorenz-96 model is a reduced-order model to describe the large-scale behaviour of the mid-latitude atmosphere \cite{lorenz96}. It consists of a set of coupled ordinary differential equations, which represent the variation of an atmospheric quantity of interest such as temperature or vorticity, on a periodic lattice representing a latitude circle on the earth 
\begin{equation}
\frac{d}{dt} x_i(t)= \left( x_{i+1}(t) - x_{i-2}(t)\right) x_{i-1}(t) - x_i(t) + F, \;\;\;\; i=1,\ldots,D,
\end{equation}
with $\mathbf{x} = \begin{bmatrix} x_1, \dots, x_D\end{bmatrix} \in \mathbb{R}^D$. The periodic boundary conditions are $ x_1(t) = x_{D+1}(t)$. Assuming a constant external forcing $F=8$ and $D=20$, the system exhibits chaotic behaviour with six positive LEs\cite{karimi2010extensivel96}. Both the numerical solution and the reference LEs are computed with the fourth-order Runge-Kutta method with a time step of $\Delta t= 0.01$. The largest Lyapunov exponent is $\lambda_1\approx 1.55$. The Kaplan-Yorke dimension is equal to $D_{KY}=13.4$. The training set consists of $N_t = 16000$ points, which is equivalent to $250 \tau_{\lambda}$. The hyperparameters can be found \textcolor{black}{in Table~\ref{table:l96_hyperparameters_lstm} the Appendix~\ref{subsec:appendix_hyperparameters}}. We deploy the PI-LSTM to reconstruct the full state in three test cases: reconstruction of (i) $D_{\xi} = 2$, (ii) $D_{\xi} = 6$, (iii) $D_{\xi} = 10 $ variables. We choose test cases (i) and (ii) to highlight the capabilities of the PI-LSTM and select (iii) to demonstrate how the network performs with significantly fewer measured variables for training than the Kaplan-Yorke dimension. When the full state is available for training, both LH-LSTM and PI-LSTM perform equally well in predicting the long-term statistics and the LEs \textcolor{black}{in the closed-loop mode}, but the PI-LSTM performs well when the training contains only partial observations. \textcolor{black}{To assess the networks' short-term forecasting capabilities, we compute the mean and standard deviation of the prediction horizon for $100$ time windows sampled from the test set. These time windows are used as warm-up input to the network, after which it evolves autonomously. Fig.~\ref{fig:l96_testcase_predictionhorizon} shows that in all three cases, the PI-LSTM achieves a larger mean prediction horizon, and therefore predicts the correct solution accurately for longer.}
\begin{figure*}[t]
    \centering
    \includegraphics[width=1\textwidth]{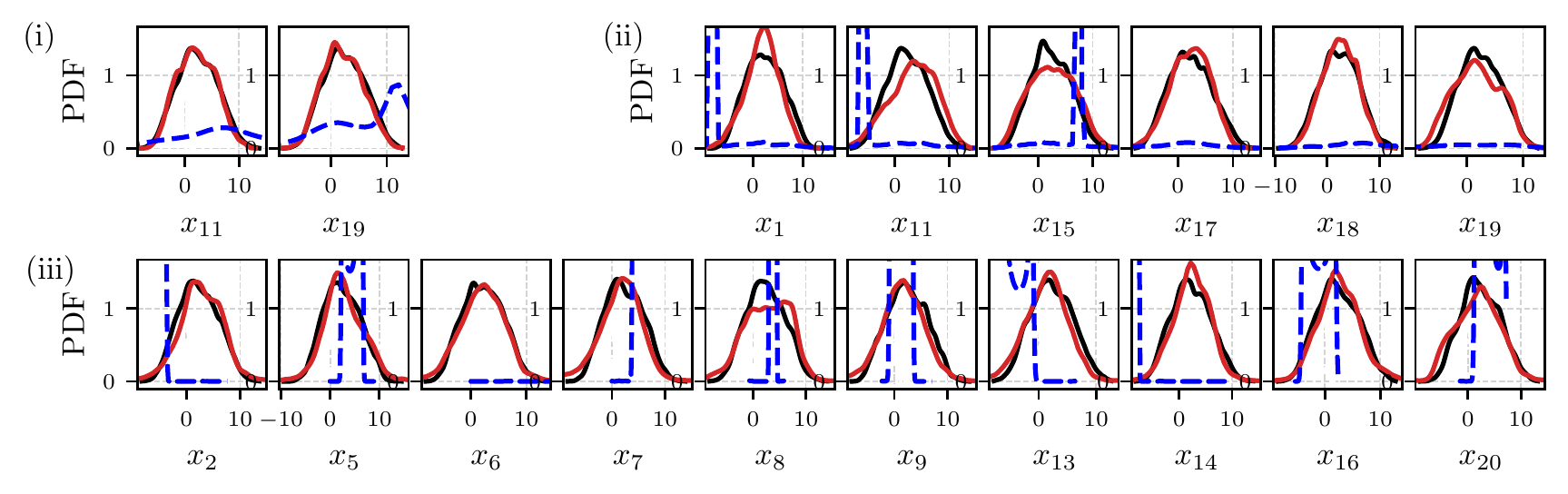}
    \caption{Lorenz-96: Statistics of reconstructed variables. Comparison of the target
(black line), PI-LSTM (red line) and LH-LSTM (blue line) probability density functions (PDF) of (i) $D_{\xi} = 2$, (ii) $D_{\xi} = 6$, (iii) $D_{\xi} = 10$ variables over a $1000\tau_{\lambda}$ trajectory in the closed-loop configuration.}
    \label{fig:l96_statistics}
\end{figure*}

The statistical reconstruction is based on an autonomous $1000\tau_{\lambda}$ long trajectory in the closed-loop mode. In Fig.~\ref{fig:l96_statistics}, we show the statistics of the hidden variables, i.e.~those variables that are not used in the training. Despite training with the full label, the LH-LSTM fails to predict the solution. In particular, the corresponding delta-like or uniform distributions (in blue) indicate a fixed-point or periodic behaviour. However, the PI-LSTM (in red) provides a markedly more accurate statistical reconstruction of the target compared to the LH-LSTM, with an average \textcolor{black}{Wasserstein distance \cite{vallender1974calculation} } of (i) $ 0.05$, (ii) $0.15$, (iii) $0.31$. 

\begin{figure}[ht]
    \centering
    \includegraphics[width=0.33\textwidth]{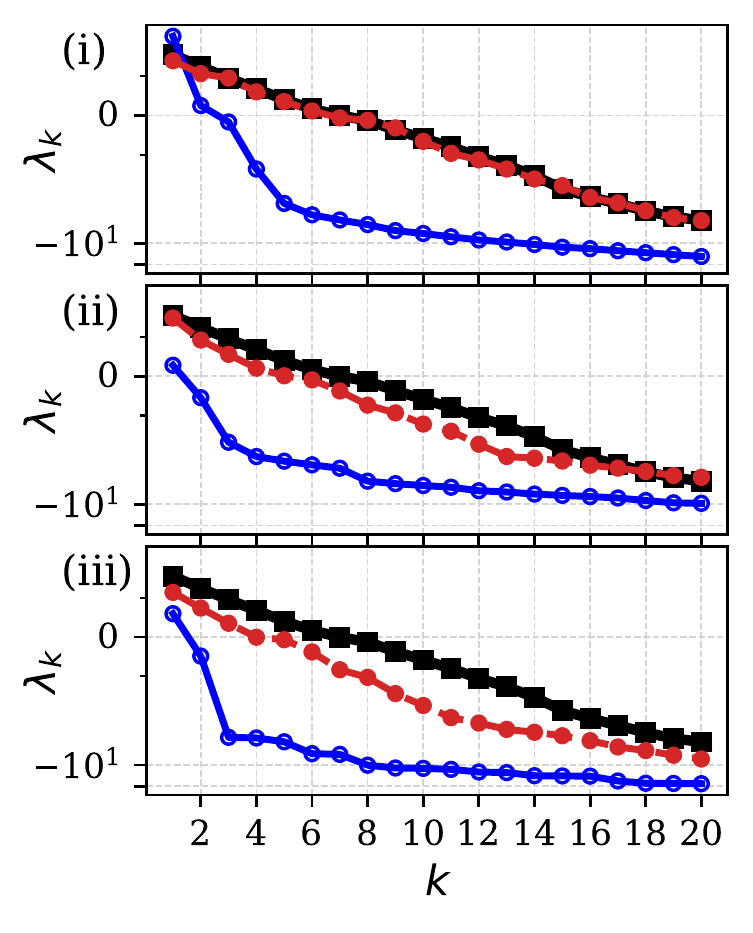}
    \caption{Lorenz-96: Comparison of the target (black squares), PI-LSTM (red dots) and LH-LSTM (blue dots) LEs for (i) $D_{\xi} = 2$, (ii) $D_{\xi} = 6$, (iii) $D_{\xi} = 10$ unobserved (hidden) variables.}
    \label{fig:l96_lyap}
\end{figure}

In Fig.~\ref{fig:l96_lyap}, we compare the reference LEs (in black squares) with LEs extracted from the data-driven LSTM (in blue circles) and PI-LSTM (in red circles) in the three test cases, \textcolor{black}{computed with Algorithm~\ref{alg:lstm_le} and hyperparameters in Table~\ref{table:l96_hyperparameters_les}}. By reconstructing the variables the networks effectively reconstruct the tangent space, the properties of which are encapsulated
in the LEs. In all cases, the LEs of the data-driven LSTM deviate significantly from the reference LEs, differing more from the target when fewer observations are available. In test case (i), the PI-LSTM correctly infers the LEs, with six positive LEs. In case (ii), the PI-LSTM infers five positive LEs, decreasing to three positive LEs for test case (iii). These findings indicate that the Kaplan-Yorke dimension of the target chaotic system (here $D_{KY}=13.4$) plays an important role in the reconstruction capabilities of PI-LSTM: as a practical rule of thumb, at least $M\gtrsim D_{KY}$ independent time series are required for a sufficient reconstruction of the full chaotic attractor, such that accurate LEs statistics are extracted. This criterion is met in cases (i) and (ii), as well as in the example of the KS equation in Sec.~\ref{sec:KS}. Still, the PI-LSTM can sufficiently reconstruct the hidden variables statistics, as seen in Fig.~\ref{fig:l96_statistics}(iii), when the number of available time series is less than $D_{KY}$.

\subsubsection{\textcolor{black}{Robustness against noise}}
\begin{figure}[h]
    \centering
    \includegraphics[width=0.5\textwidth]{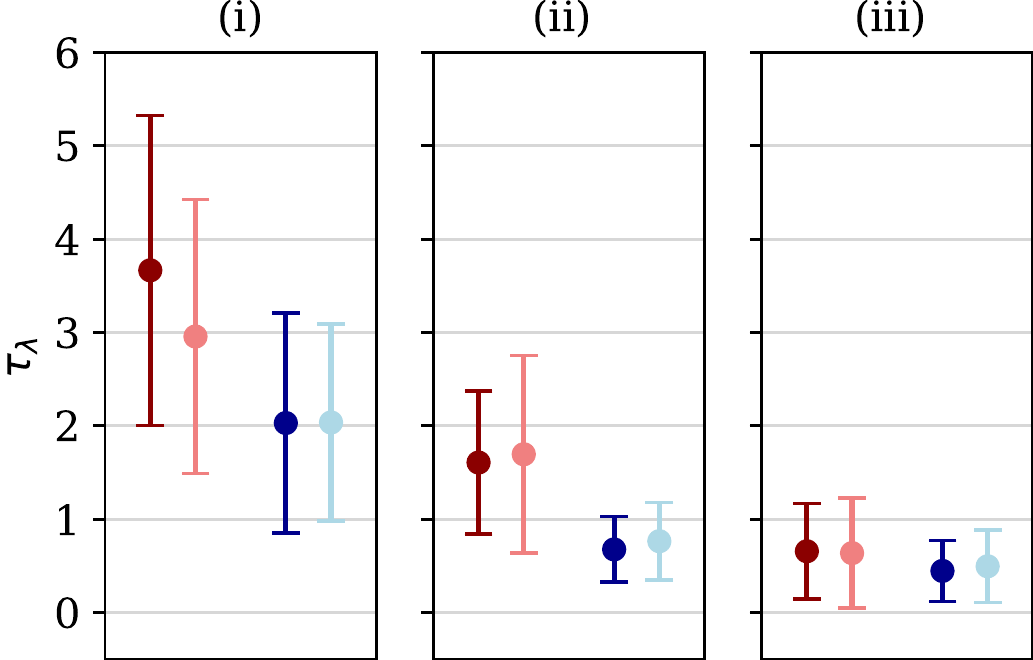}
     \caption[Lorenz-96: Prediction Horizon]{\textcolor{black}{Mean and standard deviation of the prediction horizon for  PI-LSTM (red line) and LH-LSTM (blue line) for $100$ predictions in the test set with (i) $D_{\xi} = 2$, (ii) $D_{\xi} = 6$, (iii) $D_{\xi} = 10$ missing variables and two different noise levels, $k_n=0.1$ (dark shade) and $k_n=0.2$ (light shade).} }
    \label{fig:l96_ph_noise}
\end{figure}

\begin{figure}[h]
    \centering
    \includegraphics[width=0.33\textwidth]{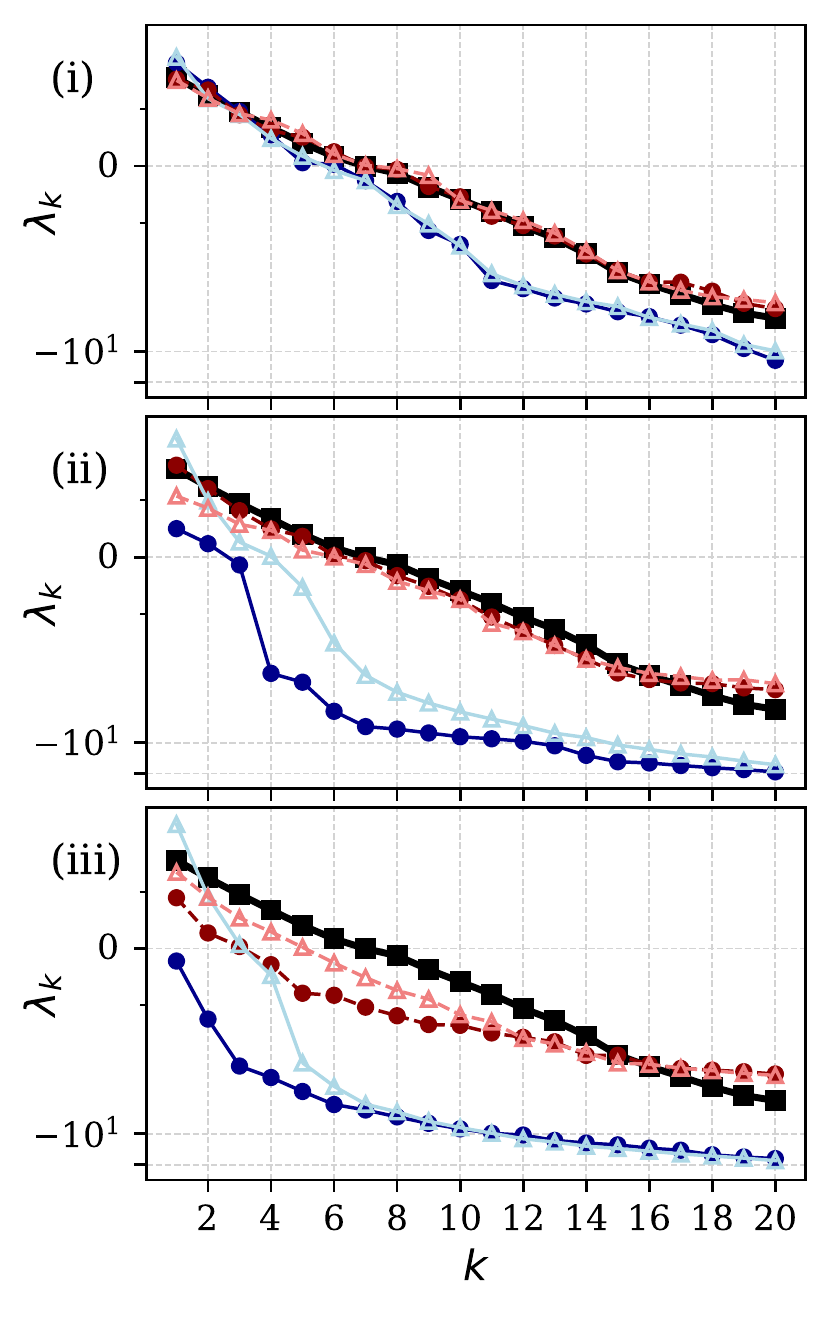}
    \caption{\textcolor{black}{ Lorenz-96: Comparison of the target (black squares), PI-LSTM (red) and LH-LSTM (blue) LEs for (i) $D_{\xi} = 2$, (ii) $D_{\xi} = 6$, (iii) $D_{\xi} = 10$ unobserved (hidden) variables for two different noise levels, $k_n=0.1$ (dark shade) and $k_n=0.2$ (light shade).}}
    \label{fig:l96_lyap_noise}
\end{figure}

\textcolor{black}{To evaluate the robustness of the networks, we examine the impact of noise on the training data to mimic experimental measurements, which may be affected by aleatoric uncertainty. Gaussian-distributed noise, sampled from $\mathcal{N}(0, k_n \sigma_{\bm{x}})$, is added onto the training data, where $\sigma_{\bm{x}}$ is the standard deviation of the data, and $k_n$ is the noise level. Subsequently, the LH-LSTM and PI-LSTM models are tested on a time window of noisy data, and their prediction horizons are computed with respect to noise-free data.
In Fig.~\ref{fig:l96_ph_noise}, we show the mean and standard deviation of short-term predictions made by the networks trained with two different noise levels, $k_n=0.1$ and $k_n=0.2$. Despite the introduction of additional noise, the models' prediction horizons remain largely consistent, which signifies that the method is robust. The PI-LSTM consistently outperforms the LH-LSTM in all three test cases, highlighting the advantages of utilizing physics-informed regularization for noisy input.
A similar observation can be made with respect to the Lyapunov spectra, as shown in Fig.~\ref{fig:l96_lyap_noise}, for which the PI-LSTM Lyapunov spectra are accurate. The probability density functions of the time series can be found in Fig.~\ref{fig:l96_statistics_n01} and Fig.~\ref{fig:l96_statistics_n02} of the Appendix~\ref{subsec:appendix_pdf_l96}. }

\FloatBarrier
\subsubsection{Discussion on the traditional methods}
As a benchmark, we reviewed a selection of methods for estimation of the Lyapunov spectrum based on time-delay embeddings \cite{WOLF1985285, ROSENSTEIN1993117,eckmann1986liapunov}. Extracting LEs from time series from high dimensional chaotic attractors is a significant numerical challenge, and it is estimated\cite{Eckmann_Ruelle_1992} that the number of required data points, $N_d$, grows exponentially with the attractor dimension $D_{KY}$ as $N_d \sim {\rm const}^{D_{KY}}$, where $\rm const \sim \mathcal{O}(10)$ encapsulates the necessary parameters of those methods. As demonstrated in \cite{Parlitz2016_EstimateLyaptimeseries} for Lorenz-96 at $D=6$ with $D_{KY}=4.2$, additional care should be given to the chosen parameters, requiring further experimentation. Based on these estimates, $N_d \geq 10^{14}$ samples are required for an adequate estimation of at least the leading Lyapunov exponent in the case of Lorenz-96 at $D=20$ used here. However, our approach achieves an accurate estimate of a large portion of the Lyapunov spectrum with significantly fewer data points ($\mathcal{O}(10^5)$).

\section{Conclusion}\label{sec:conclusion}
Because of limitations with sensors, experiments typically provide information on only part of a dynamical system. 
With only partial observations available, equation-based methods may struggle to infer the full state. 
This is a particularly challenging task in chaotic systems, which are the focus of this paper because infinitesimal perturbations exponentially grow in time. 
We propose data-driven methods to infer the dynamics of unobserved (hidden) chaotic variables -- a task that is referred to as full-state reconstruction; 
time forecast the evolution of the full state after the inference of the hidden variables; 
and compute the stability properties of the reconstructed full state. 
The tasks are performed from observations (data) limited to only part of the state with long short-term memory networks (LSTMs), which are versatile gated recurrent neural networks for sequential data (such as time series). 
First, we analyse and propose architectures. 
The first architecture is the low-to-high resolution LSTM (LH-LSTM), which takes partial observations as training input, but it requires access to the full system state (full labelled dataset). In order not to rely on the full labelled data set, we design a physics-informed LSTM (PI-LSTM), which combines data with the integral evolution equations.
This allows for the use of existing numerical integration methods, such as pseudospectral time-stepping schemes. 
The proposed reformulation is beneficial when automatic differentiation with respect to time is not readily available, and is therefore not limited to recurrent neural networks. 
Second, we mathematically derive the Jacobian of the LSTM. The Jacobian is key to computing the stability properties of the chaotic attractor such as Lyapunov exponents and covariant Lyapunov vectors. 
Third, we test both the LH-LSTM and the PI-LSTM on the chaotic Kuramoto-Sivashinsky system. Our results demonstrate that both approaches correctly perform short-term prediction and energy-spectrum inference on unseen scenarios \textcolor{black}{in the closed-loop mode}. The machines correctly learn the covariant Lyapunov vectors (measured through the angles) and infer the Lyapunov spectrum of the attractor.
Fourth, we analyse the Lorenz-96 system. Using a purely data-driven method with the LH-LSTM leads to markedly inaccurate long-term statistics and unphysical Lyapunov exponents. 
On the other hand, the proposed PI-LSTM is able to infer long-term statistics and chaotic properties. 
Fifth, the PI-LSTM outperforms the LH-LSTM by successfully reconstructing the hidden chaotic dynamics when the input dimension is smaller than the Kaplan-Yorke dimension of the attractor. This is because the missing information on the attractor is indirectly embedded in the equations. 
This work opens new opportunities for inferring hidden variables and computing the stability of dynamical systems by combining prior knowledge of the equations and data. Current work is focused on transferring the methods of this paper to experimental data and larger dynamical systems. 

\begin{acknowledgments}
Elise Özalp thanks Alberto Racca for fruitful discussions on chaos and recurrent neural networks.
This research has received financial support from the ERC Starting Grant No. PhyCo 949388. 
\end{acknowledgments}

\newpage
 \renewcommand{\thesection}{A}
\section{Appendix}\label{sec:appendix}
 \renewcommand{\thesubsection}{\arabic{subsection}}
 \renewcommand{\theequation}{\thesection\arabic{equation}}
 \setcounter{equation}{0}
\subsection{Computation of Lyapunov exponents}\label{subsec:appendix_comp_le}
\begin{algorithm}
\algrenewcommand\algorithmicrequire{\textbf{Input:}}
\algrenewcommand\algorithmicensure{\textbf{Initialize:}}
\caption{Algorithm to compute the Lyapunov exponents of an LSTM }\label{alg:lstm_le}

\begin{algorithmic}
\Require
Start state $x_t$, \\
Number of time steps to compute Lyapunov times $N^{lyap}$,\\
Number of transient initial steps to skip for warm-up $N^w$,\\
Number of steps until QR decomposition is performed again $N^{norm}$ 
\vspace{3mm}

\Ensure
\State $U \gets \textit{random}\in \mathbb{R}^{(N_h+N_h)\times d}$ \Comment{Initialize $d$ Gram Schmidt Vectors}
\State $Q, R \gets QR(U)$ \Comment{Orthonormalize GSVs}
\State $U \gets Q$
\State $N^{QR} \gets (N^{lyap} - N^w)/N^{Norm}$
\State $\lambda \gets 0\in \mathbb{R}^{d\times N^{QR}}$
\vspace{3mm}

\noindent \textit{Evolve the LSTM with its cell and hidden state and GSV simultaneously for $N^w$ steps.}
\vspace{3mm}

\For{$i = N^w: N^{lyap}$}

   \State $ x(t_{i+1}), c_{i+1}, h_{i+1} = LSTM(x(t_{i}), c_{i}, h_{i}) $ \Comment{Next LSTM step} 
   \State $J \gets Jac_{LSTM}(c_{i+1}, h_{i+1})$ \Comment{Update Jacobian}
   \State $U \gets JU $ \Comment{The variational equation}

    \If{$mod(i, N^{norm})=0$} \Comment{Orthonormalize every $N^{norm}$ steps}
    \State $Q, R \gets QR(U) $ 
    \State $U \gets Q $
    \State $\lambda[:, i/N^{norm}] \gets log(diag(R))/dt$ 
    \EndIf
\EndFor
\vspace{3mm}

\State \textbf{Output:} $\lambda_j \gets \sum_{i=0}^{N_{QR}} \lambda[j, i]/T_{lyap}$ \Comment{$j$-th Lyapunov exponent}
\end{algorithmic}

\end{algorithm}

\subsection{Computation of covariant Lyapunov vectors}\label{subsec:appendix_clvs}

Let $\bM(t, \Delta t ) = \exp\left(\int_t^{t+\Delta t} \bJ(x(\tau)) d\tau\right)$ (the exponential should be considered as a path-ordered exponential~\cite{magri2023linear}) be the system's tangent evolution operator. Each bounded non-zero CLV $\bv_i$ is evolved by the tangent dynamics $\bJ(x(t))$~\cite{huhn2020stability}
\begin{equation}
    \frac{d \bv_i}{dt} = \bJ(x(t))\bv_i - \lambda_i \bv_i,
\end{equation}
where the extra term $-\lambda_i \bv_i$ ensures that the norm is bounded. The vectors $\bv_i$ are covariant, meaning the ith CLV at time $t_1$, $\bv_i(t_1)$, maps to the ith CLV at time $t_2$, $\bv_i(t_2)$ and vice versa. This translates to $\bM(t, \Delta t)\bv_i(t) = \bv_i(t+\Delta t)$ and time-invariance follows directly as $\bM(t, -\Delta t)\bv_i(t+\Delta t) = \bv_i(t)$. 
The CLVs can be recovered from the local growth rates $\bR(t)$ and the GSVs, $\bQ(t)$, by evolving backwards in time after the forward-in-time simulation is completed. We provide a brief overview on the computation of the CLVs, for further details on the computation, we refer the reader to \cite{Margazoglou2023, Ginelli2013_clv}.
Recall that the GSV $\bQ(t)$ and growth rates $\bR(t)$ are computed by solving \eqref{autonom_dyn_system} and \eqref{tangent_equation_matrix} simultaneously with periodic orthonormalization.
After $\Delta t$, the GSV evolution is given by
\begin{align}
    \bM(t, \Delta t)\bQ(t) = \bQ(t+\Delta t)\bR(t, \Delta t)
\end{align} and the corresponding CLVs $\bV(t)$ can be written as
\begin{align}\label{GSV_CLV}
    \bV(t) = \bQ(t)\bC(t), 
\end{align}
where $\bC(t)$ is an upper triangular matrix containing the CLV expansion coefficients. From the CLV evolution equation
\begin{align}
    \bM(t, \Delta t) \bV(t) = \bV(t+\Delta t)\bD(t, \Delta t),
\end{align} we replace $\bV(t)$ using \eqref{GSV_CLV} and then rearrange. It then follows that the CLV expansion coefficients can be computed by inverting the upper triangular matrix $\bR(t, \Delta t)$  as 
\begin{align}
    \bC(t) &= \bR^{-1}(t, \Delta t)\bC(t+\Delta t)\bD(t, \Delta t),
\end{align}
after the forward-in-time computation of the GSVs is completed. This equation is evolved backwards in time starting from the end of the forward-in-time simulation. We employ the \texttt{solve\_triangular} routine of \textcolor{black}{scipy \cite{2020SciPy-NMeth}} to invert $\bR(t,\Delta t)$ and solve with respect to $\bC(t)$. The diagonal matrix $\bD(t, \Delta t)$ contains the CLV local growth factors, similar to $\bR(t,\Delta t)$. The $\bC$ and $\bD$ matrices are initialized to the identity matrix $\mathbb{I}$. We leave a sufficient spin-up and spin-down transient time at the beginning and end of our total time window, before we compute the CLVs via Eq.~\eqref{GSV_CLV}, to ensure that they are converged.

\subsection{Hyperparameters}
\label{subsec:appendix_hyperparameters}
All experiments were run on a single NVIDIA Quadro RTX 8000.
\begin{table}[h]
\begin{tabular}{ p{5cm}  p{5cm} } 
  \hline
  Hidden and cell state $N^h$ & $200, 500$\\ 
  \hline
  Batch size & $128$ \\ 
  \hline
   Physics-informed weighing $\alpha_{pi}$ &\textcolor{black}{ $1, 10, 100$ (PI-LSTM) } \newline \textcolor{black}{ $0$  (LH-LSTM)}\\
  \hline
  Tikhonov regularization &  $10^{-8}$, $10^{-9}$ \\
  \hline
  Window size &  $25$ \\ 
  \hline
  Epochs (early stopping) &  $2000$ \\ 
  \hline
  \textcolor{black}{Learning rate (Adam optimizer)} &    \textcolor{black}{$0.001$ } \\
  \hline
\end{tabular}\caption{Kuramoto–Sivashinsky equation: LSTM Hyperparameters}\label{table:ks_hyperparameters}  
\end{table}
\begin{table}[h]
\begin{tabular}{ p{5cm}  p{5cm} } 
  \hline
  Hidden and cell state $N^h$ & $100, 500$\\ 
  \hline
  Batch size & $128$ \\ 
  \hline
   Physics-informed weighing $\alpha_{pi}$ &\textcolor{black}{ $ 1, 10, 100$ (PI-LSTM) } \newline \textcolor{black}{ $0$  (LH-LSTM)}\\
  \hline
  Tikhonov regularization &  $10^{-8}$, $10^{-9}$ \\
  \hline
  Window size &  $20, 50$ \\ 
    \hline
  Epochs (early stopping) &  $2000$ \\ 
  \hline
  \textcolor{black}{Learning rate (Adam optimizer)} &    \textcolor{black}{$0.001$ }\\ 
  \hline
\end{tabular}\caption{Lorenz-96 system: LSTM Hyperparameters}\label{table:l96_hyperparameters_lstm} 
\end{table}

\begin{table}[h]
\begin{tabular}{ p{2cm}  p{2cm} } 
  \hline
  $N^{lyap}$ & $10000$ $\tau_{\lambda}$ \\ 
  \hline
  $N^{w}$ & $100$ $\tau_{\lambda}$ \\ 
  \hline
   $N^{norm}$ & $0.2$ $\tau_{\lambda}$ \\
  \hline
\end{tabular}\caption{Kuramoto–Sivashinsky equation: Algorithm~\ref{alg:lstm_le} input parameters (LEs and CLVs)} \label{table:ks_hyperparameters_les} 
\end{table}

\begin{table}[h]
\begin{tabular}{ p{2cm}  p{2cm} } 
  \hline
  $N^{lyap}$ & $100$ $\tau_{\lambda}$ \\ 
  \hline
  $N^{w}$ & $10$ $\tau_{\lambda}$ \\ 
  \hline
   $N^{norm}$ & $0.015$ $\tau_{\lambda}$ \\
  \hline
\end{tabular}\caption{Lorenz-96 system: Algorithm~\ref{alg:lstm_le} input parameters (LEs)} \label{table:l96_hyperparameters_les} 
\end{table}
\FloatBarrier
\subsection{Probability density functions of Lorenz-96 for different noise levels}\label{subsec:appendix_pdf_l96}
\FloatBarrier
\begin{figure}[h]
    \centering
    \includegraphics[width=\textwidth]{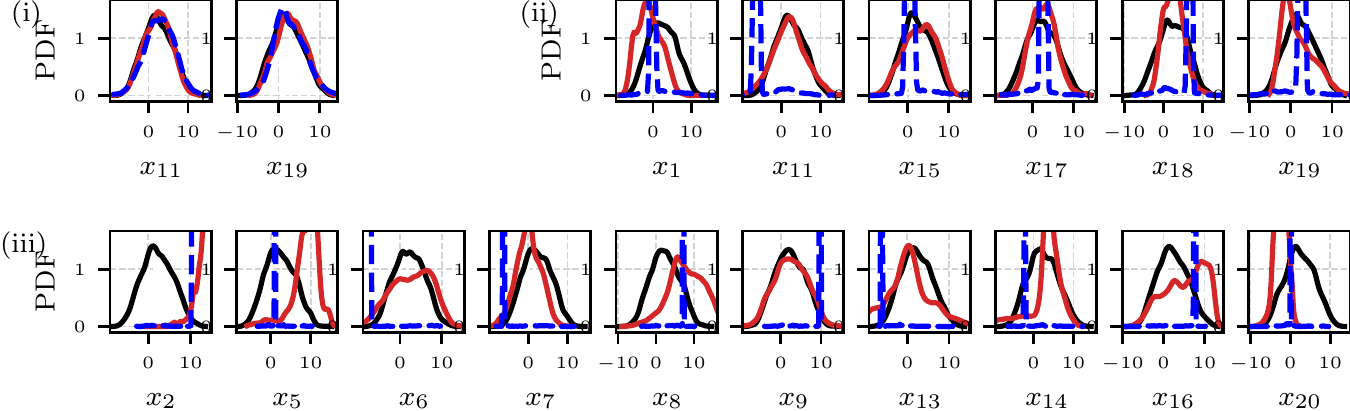}
    \caption{Lorenz-96: Statistics of reconstructed variables for $k_n = 0.1$.  Comparison of the target
(black line), PI-LSTM (red line) and LH-LSTM (blue line) probability density functions (PDF) of (i) $D_{\xi} = 2$, (ii) $D_{\xi} = 6$, (iii) $D_{\xi} = 10$ variables over a $1000\tau_{\lambda}$ trajectory in the closed-loop configuration.}
    \label{fig:l96_statistics_n01}
\end{figure}
\begin{figure}[h]
    \centering
    \includegraphics[width=\textwidth]{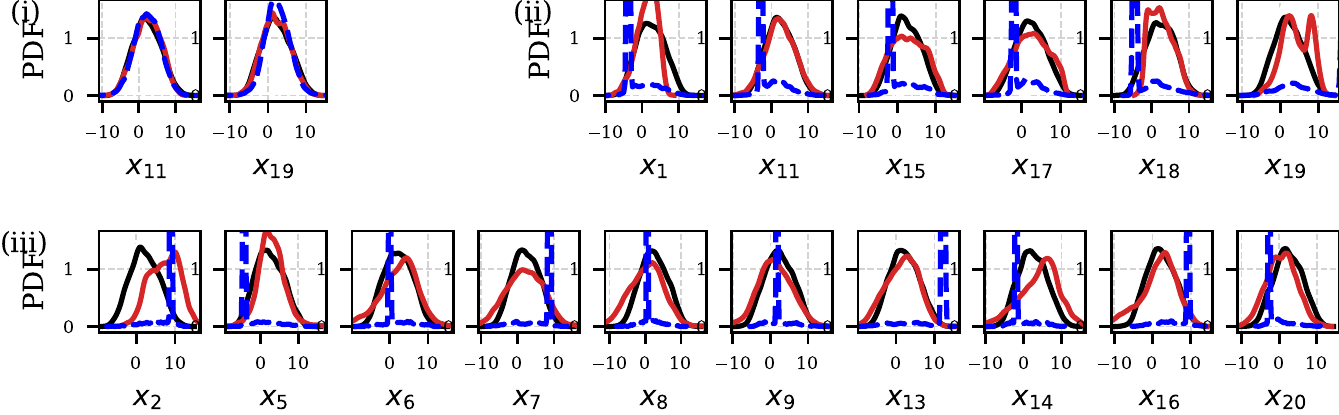}
    \caption{Lorenz-96: Statistics of reconstructed variables for $k_n = 0.2$.  Comparison of the target
(black line), PI-LSTM (red line) and LH-LSTM (blue line) probability density functions (PDF) of (i) $D_{\xi} = 2$, (ii) $D_{\xi} = 6$, (iii) $D_{\xi} = 10$ variables over a $1000\tau_{\lambda}$ trajectory in the closed-loop configuration.}
    \label{fig:l96_statistics_n02}
\end{figure}

\FloatBarrier

\subsection{CLVs of the Kuramoto–Sivashinsky equation}\label{subsec:appendix_clvs_ks}
\begin{figure}[h]
    \centering
    \includegraphics[width=\textwidth]{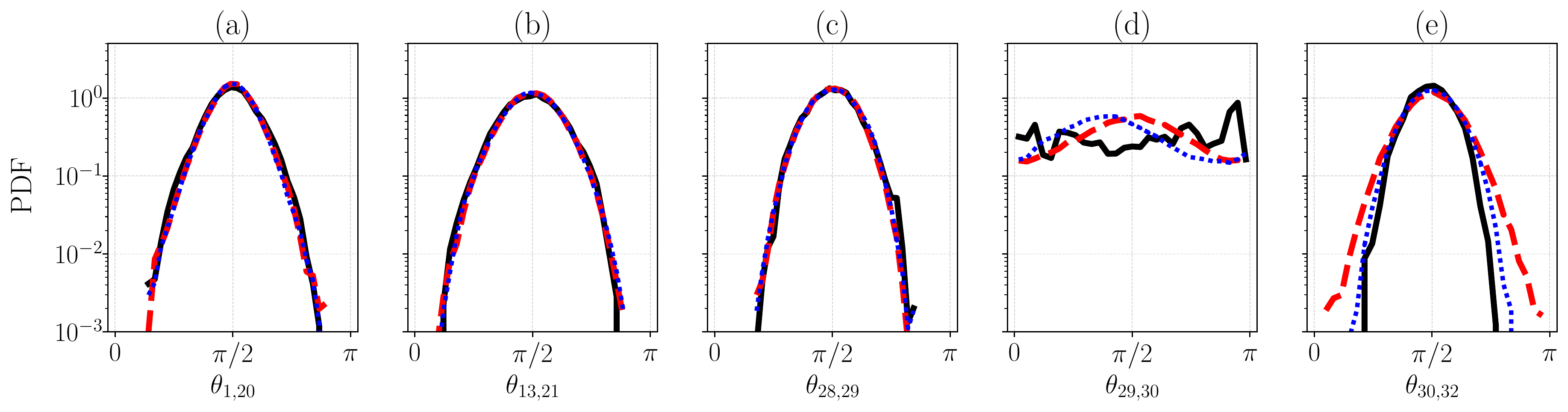}
    \caption{The angle distribution of the Kuramoto–Sivashinsky system for the CLV pairs (a) $\bv_1, \bv_{20}$, (b) $\bv_{13}, \bv_{21}$, (c) $\bv_{28}, \bv_{29}$, (d) $\bv_{29}, \bv_{30}$ and (e) $\bv_{30}, \bv_{32}$.}
    \label{fig:KS_clv_angles_appendix}
\end{figure}

In Sec.~\ref{sec:KS}, we have demonstrated that both the LH-LSTM and the PI-LSTM accurately infer the angle statistics measured between the leading CLV for the unstable $E^U_{\bm{x}}$, neutral $E^N_{\bm{x}}$, and stable $E^S_{\bm{x}}$ subspaces of the Kuramoto-Sivashinksy equation. 
Specifically, as depicted in Fig.~\ref{fig:KS_clv_angles_appendix}(a)-(c) and (e), the angle distributions exhibit a pronounced peak at $\pi/2$, then rapidly decline near $0$ and $\pi$. A change of this behaviour can be found in Fig.~\ref{fig:KS_clv_angles_appendix}(d), in which the angle distribution spans the whole $[0, \pi]$ interval. This shows the transition between physical and spurious modes, as discussed in \cite{takeuchi2011hyperbolic}. 

\nocite{*}
\bibliography{aipsamp}

\end{document}